\documentclass[11pt,preprint,3p]{elsarticle}
\usepackage{amssymb}

\begin{document}
\begin{frontmatter}
\title{An extension of gas-kinetic BGK Navier-Stokes
scheme to multidimensional astrophysical
magnetohydrodynamics}
\author{Chun-Lin Tian}
\ead{C.L.Tian@astro.elte.hu; cl\_tian@hotmail.com}
\address{E\"{o}tv\"{o}s University, Department of Astronomy,
Budapest, Pf. 32, H-1518 Hungary}

\begin{abstract}
The multidimensional gas-kinetic scheme for the
Navier-Stokes equations under gravitational fields [J.
Comput. Phys. 226 (2007) 2003-2027] is extended to resistive
magnetic flows. The non-magnetic part of the
magnetohydrodynamics equations is calculated by a BGK solver
modified due to magnetic field. The magnetic part is treated
by the flux splitting method based gas-kinetic theory [J.
Comput. Phys. 153 (1999) 334-352 ], using a particle
distribution function constructed in the BGK solver. To
include Lorentz force effects into gas evolution stage is
very important to improve the accuracy of the scheme. For
some multidimensional problems, the deviations tangential to
the cell interface from equilibrium distribution are
essential to keep the scheme robust and accurate.
Besides implementation of a TVD time discretization scheme,
enhancing the dynamic dissipation a little bit is a simply
and efficient way to stabilize the calculation.
One-dimensional and two-dimensional shock waves tests are
calculated to validate this new scheme. A three-dimensional
turbulent magneto-convection simulation is used to show the
applicability of current scheme to complicated astrophysical
flows.
\end{abstract}

\end{frontmatter}
\section{Introduction}
Plasma is a very common phase of matter in the universe, especially
under astrophysical circumstances. For instance, star can be
regarded as very hot ball of plasma. The solar surface activities
are thought to be tightly related to the interaction between
turbulent convection and magnetic flux tubes. The accretion flows
are always magnetized. Magnetic fields are also important in the
interstellar medium of spiral, barred and irregular galaxies. These
problems usually are complicated and high-nonlinear. Along with the
fast growth of computer power, numerical computation is becoming a
more and more popular way to study them.

During the past decades, to meet the demands of industrial design
and scientific computation, a lot of efforts have been made to
develop accurate and robust numerical schemes for supersonic
flows. Various methods, such as Godnov scheme, piecewise parabolic
method (PPM)\cite{ppm1984}, total variation diminishing (TVD)
scheme\cite{Harten1983357} and gas-kinetic BGK
scheme\cite{Xu19949,xu2001289}, have been invented based on
different philosophies. When calculating the shock waves, we face
the dilemma of keeping accuracy or robustness. In order to stabilize
the calculation, we need dissipation. While an unnecessary
dissipation would reduce the accuracy of the scheme. Hence, the
capability of a shock-capturing scheme depends on how smart the
introduction of the numerical dissipation is. In the gas-kinetic BGK
scheme, this is done by particle collisions. Due to collision, the
particles will approach a thermally equilibrium state, e.g.,
Maxwellian distribution. Practice shows that this is a smart way to
control dissipation. The resulted supersonic flow solver is accurate
and robust.

At the same time, some schemes for magnetohydrodynamics
(MHD) have been developed or extended from their
hydrodynamics (HD) version. For example, the PPM has been
extended to mulitdimensional MHD by Dai and
Woodward\cite{Dai1994485}. They also proposed an approximate
MHD Riemann solver and an approach to maintain the
divergence free condition of magnetic
field\cite{dai1998331}. A second-order accurate TVD Roe-type
upwind MHD scheme and its multidimensional version were
presented in \cite{ryu951,ryu952}. Most of them are based on
characteristic decomposition. Because of the non-strictly
hyperbolicity of the MHD system, it is hard to validate the
MHD eigensystem. The gas-kinetic theory based flux splitting
method for ideal MHD developed by Xu\cite{xu1999334} is a
robust and first-order scheme. It was extended to higher
order and multidimensional by Tang and Xu\cite{Tang200069}.
Without wave decomposition, it is an efficient scheme. For
the flux vector splitting (FVS) schemes\cite{Steger1981263},
the flux function is split to $F=F^{+}+F^{-}$, where the
numerical dissipation is proportional to the numerical
resolution. An equilibrium part due to particle collisions
is introduced in \cite{xu1999334}, i.e.,
$F=(1-\alpha)F^e+\alpha(F^{+}+F^{-})$. $\alpha\in[0,1]$ is
an adjustable parameter used to control the particle
collisions. In doing so, the unnecessary numerical viscosity
is reduced.

For the BGK-Navier-Stokes (BGK-NS) solver, the particle collisions
are controlled by the solution of BGK equation\cite{bgk54} instead
of an arbitrary parameter. The reduction of dissipation only takes
place where it is physically necessary. In oder to apply the BGK-NS
solver to astrophysical flows, the magnetic field has to be
included. The problem is that an implementation of the magnetic
terms with arbitrary descretization scheme might give rise to
complicated truncation error behaviour. As analyzed by
\cite{tian07}, it is better to use the same difference operator to
discretize all the terms in the equations, otherwise, the
interaction between different kinds of truncation errors would
introduce unphysical effect, e.g., \emph{numerical heating} or
\emph{numerical dispersion}. In the current study, the non-magnetic
part of MHD equations is solved by BGK-NS scheme, the magnetic part
is solved by flux splitting method based on the solution of BGK
equation.

This paper is organized as follows. Section~\ref{numsch}
describes the extension of BGK scheme to include magnetic
field. In Section~\ref{numtst} numerical experiments are
performed to validate the current scheme.
Section~\ref{discu} discusses selected problems related to
the current scheme and some general issues of developing MHD
scheme. Section~\ref{conclu} is the conclusion.

\section{Numerical Scheme}
\label{numsch}

In this section, I describe a scheme for solving the following
resistive MHD equations under gravitational field:

\begin{eqnarray}
\label{ns1}
\partial\rho/\partial t &=& -\nabla \cdot \rho \vec{v},\\
\label{ns2}
\partial\rho \vec{v}/\partial t &=& -\nabla \cdot (\rho
\vec{v}\vec{v} -\vec{B}\vec{B}) -\nabla p_{\rm
tot}+\nabla\cdot\vec{\Sigma}+\rho(-\nabla\phi),\\
\label{ns3}
\partial E/\partial t &=&-\nabla\cdot[(E+p_{\rm tot})\vec{v}
-\vec{v}\cdot\vec{\Sigma}+\vec{F}_{\rm d}
-\vec{B}\vec{B}\cdot\vec{v}-\vec{B}\times\eta
(\nabla\times\vec{B}) ]+\rho \vec{v}\cdot(-\nabla\phi),\\
\partial\vec{B}/\partial t &=& -\nabla\cdot (\vec{v}\vec{B}
-\vec{B}\vec{v}-\eta\nabla\vec{B})
\end{eqnarray}
where, $p_{\rm tot}=p_g+p_m$, $E=E_i+E_m+E_k$, $p_g$ is the gas
pressure, $p_m=\frac{1}{2}\vec{B}\cdot\vec{B}$ is the magnetic
pressure, $E_i$ is the internal energy,
$E_m=\frac{1}{2}\vec{B}\cdot\vec{B}$ is the magnetic energy and
$E_k=\frac{1}{2}\rho v^2$ is the kinetic energy. $(-\nabla\phi)$ is the
gravitational acceleration, $F_d$ the diffusive heat flux,
$\vec{\Sigma}$ the viscous stress tensor, $\eta$ the magnetic
resistivity. All the other symbols have their standard meaning.  The
above equations can be written in a compact form:
\begin{equation}
\frac{\partial U}{\partial t}=\frac{\partial F_x}{\partial x}+
\frac{\partial F_y}{\partial y}+\frac{\partial F_z}{\partial z}+Q
\end{equation}
where $U=(\rho,\rho\vec{v},\vec{B},E)^T$ and
$Q=(0,\rho(-\nabla\phi),0,\rho\vec{v}\cdot(-\nabla\phi))^T$ are cell averaged
quantities. $F_x$, $F_y$ and $F_z$ are defined on the cell
interface. In each direction, the fluxes can be split into two
parts. For example, $F_x$ can be further written as
$F_x=F_{x,\rm{bgk}}+F_{x,\rm{split}}$, where
\begin{eqnarray}
F_{x,\rm bgk}=-\left(\begin{array}{c} \rho v_x\\ \rho v_x
v_x+p_{tot}-\Sigma_x\\ \rho v_x v_y-\Sigma_y\\ \rho v_x
v_z-\Sigma_z\\ 0\\ 0\\ 0\\
 (E+p_{\rm tot})v_x-\kappa\nabla_x T-v_x \Sigma_x
\end{array}\right),
\end{eqnarray}
and
\begin{eqnarray}
F_{x,\rm split}=\left(\begin{array}{c} 0\\ B_xB_x\\
B_xB_y\\ B_xB_z\\ \eta \nabla_xB_x\\ B_xv_y-v_xB_y+\eta
\nabla_xB_y\\ B_xv_z-v_xB_z+\eta \nabla_xB_z\\
B_x(B_yv_y+B_zv_z+B_xv_x)+\eta[\vec{B}\times(\nabla\times\vec{B})]_x
\end{array}\right).
\end{eqnarray}
We discretize the non-magnetic part $F_{x,\rm bgk}$ by gas-kinetic
BGK scheme, magnetic part $F_{x,\rm split}$ by gas-kinetic theory
based flux splitting method.

\subsection{A BGK-NS solver for the non-magnetic part}

\begin{figure}
\centering
\includegraphics[scale=1]{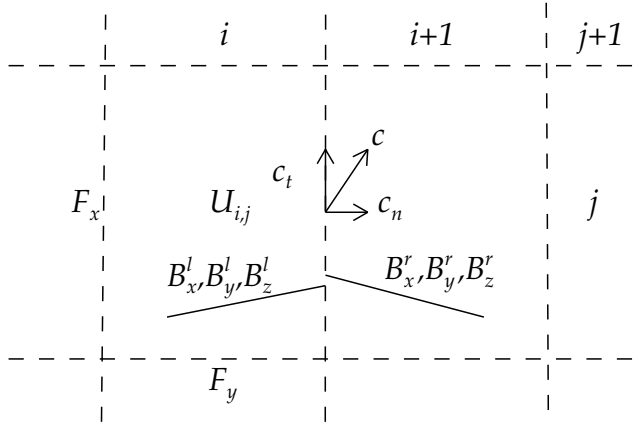}
\caption{Two adjacent finite volumes -- cells. (i,j) are the cell
indices. Dashed lines are the cell interfaces. Fluxes of the
conservative quantities, eg., $F_x$ and $F_y$ are calculated at the
cell boundaries. The cell-averaged $U_{i,j}$ is updated every time
step. Solid lines show schematically the distribution of magnetic
field constructed by van Leer limiter around cell interface if there
is a shock. Arrows are a vector and its normal and tangential
components.} \label{figcell}
\end{figure}

The current scheme calculates $F_{x,\rm{bgk}}$ by the BGK-NS solver
\cite{tian07}, which is needed to be modified to take into account
the effects of magnetic field. First, the total pressure includes
magnetic pressure, and total energy magnetic energy. Secondly, the
effects of Lorentz force are added into the BGK
equation\cite{bgk54}, i.e.,

\begin{equation}
\label{bgk} \frac{\partial f}{\partial t}+\vec{c}\cdot\nabla
{f}+(-\nabla\phi+\vec{F}_{Lorenzt})\cdot\nabla_{\vec{c}}f =\frac{g-f}{\tau},
\end{equation}
where $f(t,x,y,z,c_x,c_y,c_z,\xi)$ is the particle
distribution function defined on phase space, $g$ is the
Maxwellian distribution,
\begin{equation}
g=\rho\left(\frac{\lambda}{\pi}\right)^{\frac{N+3}{2}}e^{-\lambda((\vec{c}-
\vec{v})\cdot(\vec{c}- \vec{v})+\xi^2+\vec{B}\cdot\vec{B}/2)},
\end{equation}
$\tau$ is relaxation time, $\vec{c}=(c_x,c_y,c_z)$ is the
particle micro velocity. $\xi$ is the internal variable,
which has $N$ internal degree of freedom. The value of $N$
is determined by the configuration of molecules. $\lambda$
is related to the gas temperature by $\lambda=m/2kT$, $m$ is
the molecules mass, $k$ is the Boltzmann constant, and $T$
is the temperature.
$\vec{F}_{Lorentz}=\vec{B}\times(\nabla\times\vec{B})/\rho$
is the acceleration due to Lorentz force.
Equation~(\ref{bgk}) has a local approximate solution at
cell interface\cite{tian07}:

\begin{eqnarray}
f=&&\{(1-e^{-t/\tau})+(e^{-t/\tau}(t+\tau)-\tau)[(\bar{a}_n^lc_n+
\bar{b}_n^l({-\nabla\phi}+\vec{F}_{Lorentz})_n)H[c_n]\nonumber\\
&&+(\bar{a}_n^rc_n+\bar{b}_n^r({-\nabla\phi}+\vec{F}_{Lorentz})_n)(1-H[c_n])\nonumber\\
&&+\bar{\vec{a}}_t\cdot\vec{c}_t+
\bar{\vec{b}}_t\cdot({-\nabla\phi}+\vec{F}_{Lorentz})_t]+[t-\tau(1-e^{-t/\tau})]
\bar{A}\}g_0\nonumber\\
&+&e^{-t/\tau}\{[1-\vec{a}^l\cdot\vec{c}t-\vec{b}^l\cdot({-\nabla\phi}
+\vec{F}_{Lorentz})t\nonumber\\
&&-\tau(\vec{a}^l\cdot\vec{c}
+\vec{b}^l\cdot({-\nabla\phi}+\vec{F}_{Lorentz})+A^l)]H[c_n]g_0^l\nonumber\\
&+&[1-\vec{a}^r\cdot\vec{c}t-\vec{b}^r\cdot({-\nabla\phi}
+\vec{F}_{Lorentz})t\nonumber\\
&&-\tau(\vec{a}^r\cdot\vec{c}
+\vec{b}^r\cdot({-\nabla\phi}+\vec{F}_{Lorentz})+A^r)](1-H[c_n])g_0^r\},
\label{stbgk}
\end{eqnarray}
where the Heaviside function $H[x]$ is defined by
\begin{eqnarray*}
H[x]=\left\{
\begin{array}{l}
0, x < 0, \\ 1, x\ge 0,
\end{array}
\right.
\end{eqnarray*}
$x=0$ is the location of cell interface. For a vector defined on the
cell interface, its components are grouped into two categories, the
component normal to the cell interface and those tangential
components. For instance, for velocity $\vec{c}$, $c_n$ is the normal
component, $\vec{c_t}$ are the tangential components. See
Fig.~\ref{figcell} for an example.
$\bar{\vec{a}}^l=[\bar{a}_n^l,\bar{\vec{a}}_t^l]$,
$\vec{a}^l=[a_n^l,\vec{a}_t^l]$, $
\bar{\vec{b}}^l=[\bar{b}_n^l,\bar{\vec{b}}_t^l]$, $
\vec{b}^l=[b_n^l,\vec{b}_t^l]$,
$\bar{\vec{a}}^r=[\bar{a}_n^r,\bar{\vec{a}}_t^r]$,
$\vec{a}^r=[a_n^r,\vec{a}_t^r]$, $
\bar{\vec{b}}^r=[\bar{b}_n^r,\bar{\vec{b}}_t^r]$, $
\vec{b}^r=[b_n^r,\vec{b}_t^r]$,
 $A^l$, $A^r$ and $\bar{A}$ are the coefficients measuring
the deviations from Maxwellian due to various effects. Calculation
of these coefficients is one of the key part of BGK scheme. The
details are presented in \cite{tian07}. The way that current scheme
including magnetic field does not affect these calculations. $g_0$,
$g_0^l$, $g_0^r$ are equilibrium and split Maxwellian, respectively.
Once the solution of BGK is found, the fluxes of conservative
quantities through the cell interface can be obtained, i.e.,
\begin{eqnarray}
\label{flux1}
F_{x,\rm{bgk}}&=&-\int c_x f\vec{\psi}d\Xi,\\
\vec{\psi}&=&[1,\vec{c},0,0,0,
\frac{1}{2}(\vec{c}\cdot\vec{c}+\xi^2+\vec{B}\cdot\vec{B})]^T,
\end{eqnarray}
where $d\Xi=dc_xdc_ydc_zd\xi$ is the volume element in phase space.
The above integrals can be worked out analytically, which involves
another key part of the BGK scheme -- calculation of the moments of
Maxwellian. The formula needed can be found in the appendix of
\cite{xu1999334} or \cite{tian07}.

\subsection{A flux splitting solver for the magnetic part}
In order to perform the flux splitting, the solution of BGK equation
obtained in previous section has to be split into left-side and
right-side status, i.e.,

\begin{eqnarray}
f^l(c_n > 0)=&&\{1+(e^{-t/\tau}(t+\tau)-\tau)[(\bar{a}_n^lc_n+
\bar{b}_n^l({-\nabla\phi}+\vec{F}_{Lorentz})_n)\nonumber\\
&&+\bar{\vec{a}}_t^l\cdot\vec{c}_t+
\bar{\vec{b}}_t^l\cdot({-\nabla\phi}+\vec{F}_{Lorentz})_t]
+[t-\tau(1-e^{-t/\tau})]
\bar{A}\nonumber\\
&+&e^{-t/\tau}[\vec{a}^l\cdot\vec{c}t+\vec{b}^l\cdot({-\nabla\phi}
+\vec{F}_{Lorentz})t\nonumber\\
&&+\tau(\vec{a}^l\cdot\vec{c}
+\vec{b}^l\cdot({-\nabla\phi}+\vec{F}_{Lorentz})+A^l)]\}g_0^l,
\end{eqnarray}
\begin{eqnarray}
f^r(c_n\leq 0)=&&\{1+(e^{-t/\tau}(t+\tau)-\tau)[(\bar{a}_n^rc_n+
\bar{b}_n^r({-\nabla\phi}+\vec{F}_{Lorentz})_n)\nonumber\\
&&+\bar{\vec{a}}_t^r\cdot\vec{c}_t+
\bar{\vec{b}}_t^r\cdot({-\nabla\phi}+\vec{F}_{Lorentz})_t]
+[t-\tau(1-e^{-t/\tau})]
\bar{A}\nonumber\\
&+&e^{-t/\tau}[\vec{a}^r\cdot\vec{c}t+\vec{b}^r\cdot({-\nabla\phi}
+\vec{F}_{Lorentz})t\nonumber\\
&&+\tau(\vec{a}^r\cdot\vec{c}
+\vec{b}^r\cdot({-\nabla\phi}+\vec{F}_{Lorentz})+A^r)]\}g_0^r,
\end{eqnarray}

Two points need to be remarked on the above splitting.
\begin{enumerate}
  \item The tangential derivations due to derivatives and accelerations
    are very important for the BGK-NS solver. However, they do not
    affect the magnetic part evidently. This property seems
    case-dependent. For example, for the Orzang-Tang vortex problem,
    including tangential deviations is little bit dispersive.
     While for the cloud-shock
    interaction problem, it essentially improve the ability of the scheme.
    See Section~\ref{numtst} for the details.
  \item The right-hand side of Equation~(\ref{stbgk})
    consists three parts, i.e.,
    equilibrium part $[\cdots]g_0$, split parts $[\cdots]g_0^l$ and
    $[\cdots]g_0^r$. The left and right parts introduce
    \emph{kinematic} dissipation to stabilize calculation
    of discontinuity;
    while the equilibrium part reduces the kinematic dissipation
    to an appropriate value, thus improve the accuracy of the
    numerical scheme. These three terms together form a distribution which
    would approach an equilibrium state due to the particle collision.
    If we split the distribution (\ref{stbgk}) into three parts accordingly,
    and replace those squares with an adjustable constant
    $\alpha, (0\leq\alpha\leq 1)$,
    i.e., $(1-\alpha) g_0$, $\alpha g_0^l$ and $\alpha g_0^r$, then
    the following fluxes construction is the same as in method \cite{xu1999334}.
    However, numerical experiment shows that this kind of splitting
    seems too dispersive, even without substituting those coefficients.
    This will be discussed further in Section~\ref{discu}.
\end{enumerate}

Using the above split particle distribution, the magnetic part of
fluxes can be split as: $F_{x,\rm{split}}=F_x^{l}+F_x^{r}$.
$F_x^{l}$ is computed as follows,
\begin{eqnarray}
F_{x}^{l}=&&\frac{1}{\rho^l}\int^{\infty}_0 f^{l}_s
\left(\begin{array}{c}
0\\
B_x^lB_x^l\\
B_x^lB_y^l\\
B_x^lB_z^l\\
\eta\nabla_xB_x^l\\
B_x^lv_y^l+\eta\nabla_xB_y^l\\
B_x^lv_z^l+\eta\nabla_xB_z^l\\
B_x^l(B_y^lv_y^l+B_z^lv_z^l+\frac{1}{2}B_x^lv_x^l)
+\eta[\vec{B^l}\times(\nabla\times\vec{B^l})]_x
\end{array}\right) d\Xi \nonumber\\
&+&\frac{1}{\rho^l}\int^{\infty}_0 f^{l} c_x\left(\begin{array}{c}
0\\
0\\
0\\
0\\
0\\
-B_y^l\\
-B_z^l\\
\frac{1}{2}B_x^lB_x^l
\end{array}\right) d\Xi,
\label{flxspl}
\end{eqnarray}
where $(B_x^l,B_y^l,B_z^l)$ and $(B_x^r,B_y^r,B_z^r)$  are the
magnetic field at the left-hand side and right-hand side of the cell
interface, respectively. They are reconstructed at the beginning of each
time step at the cell interface.
\begin{equation}
\label{rc2} \vec{B}^l=\vec{B}_{i} (x_i) + (x_{i+1/2}-x_i) L_i ,
\end{equation}
\begin{equation}
\label{rc3} \vec{B}^r=\vec{B}_{i+1} (x_{i+1}) -(x_{i+1}-x_{i+1/2})
L_{i+1}.
\end{equation}
In the above expressions, the van Leer limiter is defined as,
\begin{displaymath}
L_i=S(s_+,s_-)\frac{|s_+||s_-|}{|s_+|+|s_-|},
\end{displaymath}
where $S(s_+,s_-)=\mathrm{sign}(s_+)+\mathrm{sign}(s_-)$,
$s_+=(\vec{B}_{i+1}-\vec{B}_i)/(x_{i+1}-x_i)$, and
$s_-=(\vec{B}_{i}-\vec{B}_{i-1})/(x_{i}-x_{i-1})$. $\rho^l$ is
constructed in the same way. Fig.~\ref{figcell} gives a schematic
 example for the reconstruction.

$F_x^{r}$ is calculated in the similar way, using $f^r$ and
integrating on the interval $[-\infty,0]$ . $F_y$ and $F_z$ are
obtained by properly permuting indices.

Some points need to be remarked on the above fluxes splitting,
\begin{enumerate}
 \item The advective magnetic terms are treated to be passively transported
 by the motions of particles. $v_y, v_z$ in 1D problems and $v_z$ in 2D
 problems are also passively transported.
 Those non-advective terms are weighted by the particle distribution.
 \item Basically, the integrals on the right-hand-side of (\ref{flxspl})
 are separated according to the following rule. If a term
 explicitly contains the component of velocity norm to
 the cell interface, eg., $v_x$, then $v_x$ is replaced
 by the particle velocity $c_x$
 and grouped into the second integral.
 If not, it is grouped into the first integral.
 \item An exception of the above rule is the term $v_x^lB_x^lB_x^l$.
 Because in Eq.~(\ref{ns3}),we already include the
 magnetic pressure $p_m$ and magnetic energy density $E_m$ into the total
 pressure $p_{tot}$ and total energy $E$.
 The transported magnetic contributions to $p_{tot}$ and $E$,
 eg., $v_xB_xB_x$ should be canceled by
 the terms from $\vec{B}\vec{B}\cdot\vec{v}$ in Eq.~(\ref{ns3}).
 The BGK scheme automatically divides
 $p_m$ and $E_m$ into two parts, eg., $p_m=(v_xp_m+c_xp_m)/2$.
 In order to cancel these terms,
 we split $v_x^lB_x^lB_x^l$ into:
 $f^l c_x^l B_x^l B_x^l/2$ and
 $f^l v_x^l B_x^l B_x^l/2$, accordingly.
\end{enumerate}

\subsection{Equation of State}

Internal variable $\xi$ is used to take account of the
realistic effects of molecular structure to equation of
state (EOS). For 3D diatomic molecule ideal gas,
$\rho\xi^2/2$ is the sum of vibrational and rotational
energy. $\xi$ is a function of the effective internal
freedom degree $K$. For 3D cases, $K=N$. For 2D cases,
$K=N+1$, where the particle's motion in the additional third
direction is included. For 1D problems, $K=N+2$. Instead of
temperature $T$ and internal energy $E_i$, in the gas
evolution stage of the BGK scheme, $\lambda$ and internal
freedom degree $K$ are involved explicitly. For non-magnetic
ideal gas, they are defined as:

\begin{eqnarray}
 \lambda&=&\frac{\rho}{2p},\\
 K&=&4(E_{i})\lambda/\rho-3.
\end{eqnarray}
For general cases, we define effective $\lambda^{\ast}$ and
$K^{\ast}$ as following:
\begin{eqnarray}
 \lambda^{\ast}&=&\frac{\rho}{2p_{tot}},\\
 K^{\ast}&=&4(E_{tot}-E_k)\lambda^\ast/\rho-3,
\end{eqnarray}
where, $p_{tot}=p_g+p_r+p_m$ and $E_{tot}=E_i+E_r+E_m+E_k$. $p_m$
and $E_m$ are the pressure and energy due to magnetic field. $p_r$
and $E_r$ can be the pressure and energy due to radiation or
ionization.

\subsection{Controlling Dissipation}
Besides MUSCL type of reconstruction, the numerical dissipation can
be determined explicitly. The aforementioned collision time is
defined as following,
\begin{equation}
\label{vis}\tau=C_1\Delta
t+C_2\frac{|p^l_{tot}-p^r_{tot}|}{|p^l_{tot}+p^r_{tot}|}\Delta
t+\frac{\mu}{p_{tot}},
\end{equation}
where $\Delta t$ is numerical time-step, $\mu$ is the dynamic
viscosity, $p^l_{tot}$, $p^r_{tot}$ are reconstructed total pressure
at the left- and right-side of the cell interface. $C_2=1$. For viscous
flows, $C_1=0$, for inviscid flows, $C_1=0.1\sim 0.2$.  For magnetic
flows, usually, the resistivity $\eta$ is set to zero for ideal MHD.
Sometimes, a small value of resistivity, e.g., $\eta=0.008\Delta t$
is needed to stabilize the calculation.

\subsection{Ensuring $\nabla\cdot\vec{B}=0$}
The divergence free condition of magnetic field means the
nonexistence of magnetic monopoles. Although in theory, they
possibly exist, in reality they have never been found. On the
contrary, caused by numerical truncation error, in MHD simulation,
non-zero $\nabla\cdot\vec{B}$ is accumulated rapidly during time
marching and destroys the physical consistence of the system. In
practice, there are different ways to suppress such kind of
accumulation.
 A detailed comparison of these method is given by T\'{o}th
\cite{toth2000605}. In 8-wave scheme, the non-zero divergence is
propagated away. More often, people use projection method to correct
$\vec{B}$ obtained from MHD scheme at time-step $n+1$. Through
$\vec{B}^\ast=\vec{B}^{n+1}+\nabla\Theta$, where
$\nabla^2\Theta+\nabla\cdot\vec{B}^{n+1}=0$,
$\nabla\cdot\vec{B}^\ast=0$ is satisfied. Since this method needs
solving a Possion equation, it is not very efficient for 3D problems
solved on very large grids. In the current study, I use projection
method for 2D programs. For 3D programs, I employ the so-called
field-interpolated constrained transport scheme \cite{dai1998331} to
ensure the divergence free constraint. In this method, perfect
cancellation of terms makes the divergence unchanging during time
marching. It is simply and time saving. After obtaining the
cell-averaged $\vec{B}$ at time $t^{n+1}$ by gas kinetic scheme, we
have magnetic field at time $t^{n+1/2}$,
$\vec{B}^{n+1/2}=0.5(\vec{B}^{n}+\vec{B}^{n+1})$. $\vec{B}^{n+1/2}$
is used to evaluate the right-hand-side of Eq.~(\ref{induc}).
\begin{equation}
\label{induc}
\frac{\partial \vec{B}}{\partial t}=\nabla\times
(\vec{v}\times\vec{B}-\eta\nabla\times\vec{B}).
\end{equation}
The left-hand-side of Eq.~{(\ref{induc})} represents the values on
the staggered mesh. After the values on staggered mesh are updated,
the new cell-averaged values are obtained though interpolation. In
principle, this method can ensure divergence free constraint to
machine error. However, the truncation error occurring in boundary
ghost cells propagates into the computational domain. Finally, we
maintain $\nabla\cdot\vec{B}=0$ to the truncation error.

After the correction of the magnetic field, the total energy should
be modified also to make the computation physically consistent, i.e.,
\begin{equation}
E^{n+1}=E^{n+1}-\frac{1}{2}(\vec{B}^{n+1}\cdot\vec{B}^{n+1})+
\frac{1}{2}(\vec{B}^{\ast}\cdot\vec{B}^{\ast})
\end{equation}

\section{Numerical Tests}
\label{numtst}

For all the tests presented in this section, the van Leer limiter is
used for the initial reconstruction of the conservative variables.
In order to make the comparison as precise as possible, for some
problems, a constant numerical time-step is used.

\subsection{1D MHD Shock Waves}
\begin{figure}
\centering
\includegraphics[scale=1]{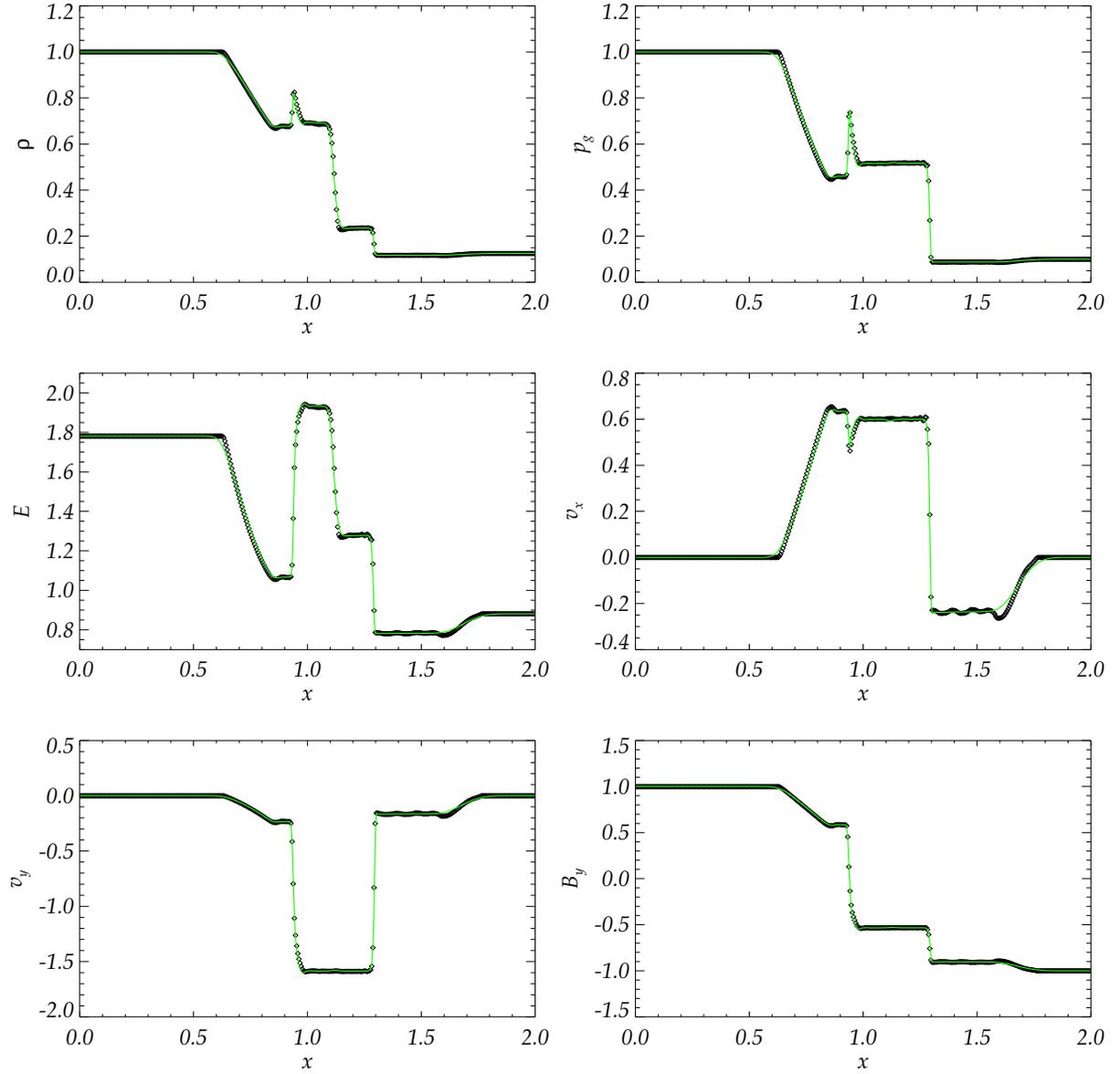}
\caption{The MHD shock tube test. A second order TVD Runge-Kutta
scheme is used for the time stepping. Dots are the result of
the Maxwellian based gas-kinetic flux splitting scheme with
$\alpha=0.9$. Lines are the result of the current MHD BGK
scheme. The simulations use 400 cells and a time step of
$\Delta t=0.2\Delta x$, corresponding to a Courant constant of $\sim
0.78$. A ratio of specific heats $\gamma=2$ is adopted. The
plotted quantities are density $\rho$, gas pressure $p_g$,
total energy density $E$, x-velocity $v_x$, y-velocity $v_y$
and y-magnetic field $B_y$ at $t=0.2$.} \label{figcop1}
\end{figure}
\begin{figure}
\centering
\includegraphics[scale=1]{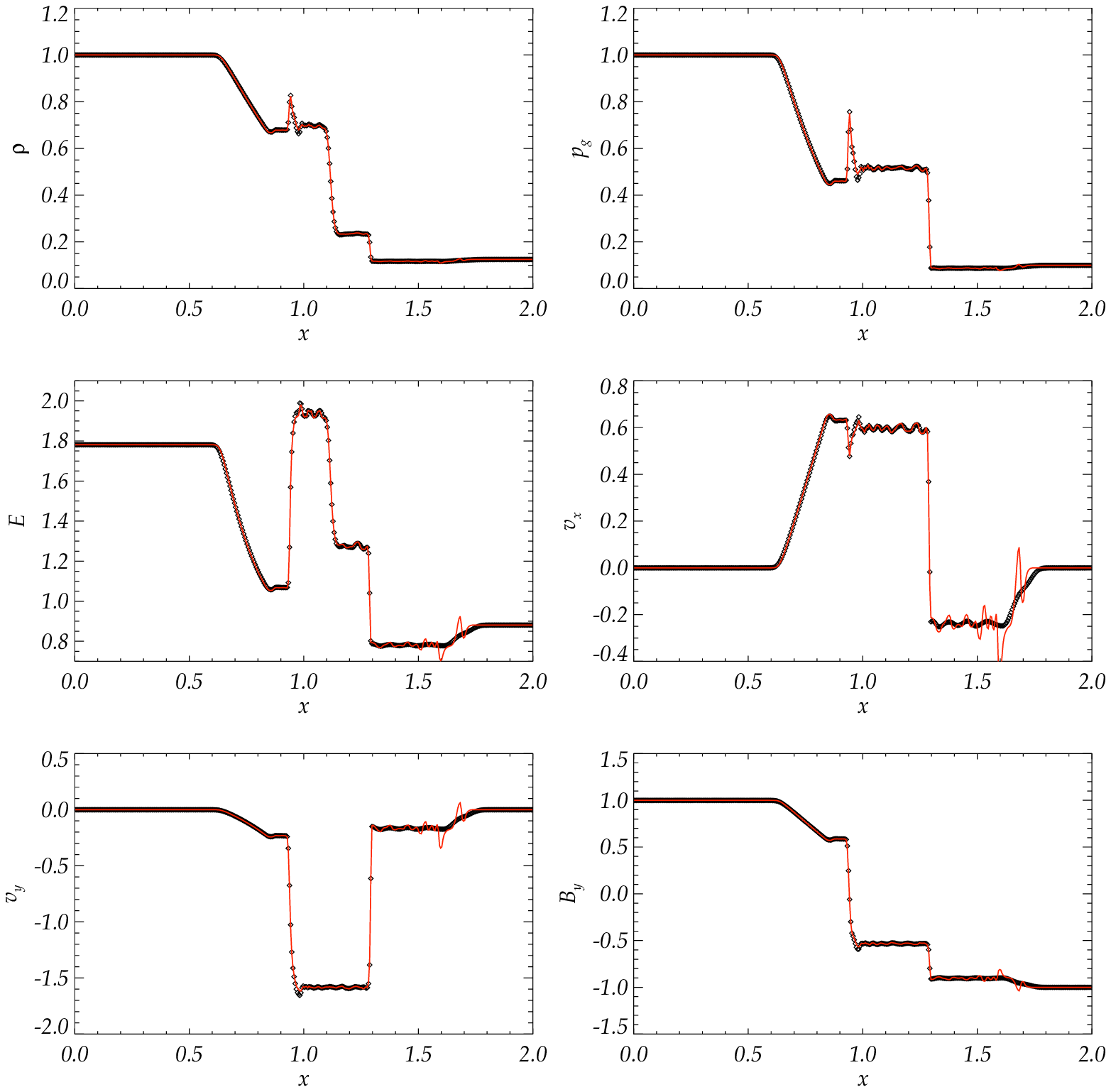}
\caption{The MHD shock tube test. The Euler time-stepping scheme is
used. Dots are the result of the current MHD BGK scheme with
effects of Lorentz force included in the gas evolution
stage. Lines are the result of the MHD BGK scheme without
considering Lorentz force in the particle distribution function.
The simulations use 400 cells and a time step of $\Delta
t=0.2\Delta x$, corresponding to a Courant constant of 0.78.
A ratio of specific heats $\gamma=2$ is adopted. The plotted
quantities are density $\rho$, gas pressure $p_g$, total
energy density $E$, x-velocity $v_x$, y-velocity $v_y$ and
y-magnetic field $B_y$ at $t=0.2$.} \label{figcop2}
\end{figure}
To validate the scheme described in Section~\ref{numsch} and
check the effects of Lorentz force in the gas evolution stage,
I test the Biro-Wu 1D MHD shock problem\cite{Brio1988400}.
The calculations are performed on $x\in[0,2]$. The initial
conditions are set up with the left state $(\rho, v_x, v_y,
v_z, p_g, B_x, B_y, B_z) =(1,0,0,0,1,0.75,1,0)$ and the
right state $(0.125,0,0,0,0.1,0.75,-1,0)$. Density $\rho$,
gas pressure $p_g$, total energy density $E$, x-velocity
$v_x$, y-velocity $v_y$ and y-magnetic field $B_y$ at time
$t=0.2$ are plotted in Fig.~\ref{figcop1} and
Fig.~\ref{figcop2}. When computing the solutions in
Fig.~\ref{figcop1}, a second-order TVD Runge-Kutta
scheme\cite{shu88} is used to do time-marching. The
solutions in Fig.~\ref{figcop2} are obtained by Euler
time-marching scheme. Fig.~\ref{figcop1} compares the gas
evolutions govern by the Maxwillian or the solution of BGK
equation. We can see the stability is increased by using
solution of BGK equation. Fig.~\ref{figcop2} shows the
effect of including the Lorentz force in gas evolution
stage. The oscillations between the front of shock and the
front of fast wave are greatly suppressed. This kind of
suppressing is not obvious in second order time marching
scheme, which means that the Lorentz force plays a role of
first order time accuracy effects in the gas evolution.

\subsection{Spherical Explosion\cite{zachary263}}
This test is used to test the influence of strong magnetic field to
the shock wave propagation. The computational domain is
$[0,1]\times[0,1]$. Initially, the density is 1 everywhere. There is
a high pressure region  around the center with a radius of $0.1$.
The pressure inside and outside the central region are 100 and 1,
respectively. The magnetic field is initialized as $(B_x, B_y,
B_z)=(0, 50/\sqrt{\pi} ,0)$. The ratio of specific heats $\gamma$
equals 2. A uniform $100\times 100$ mesh is used for this problem.
The boundaries are periodic.
\begin{figure}
\centering
\includegraphics[scale=0.9]{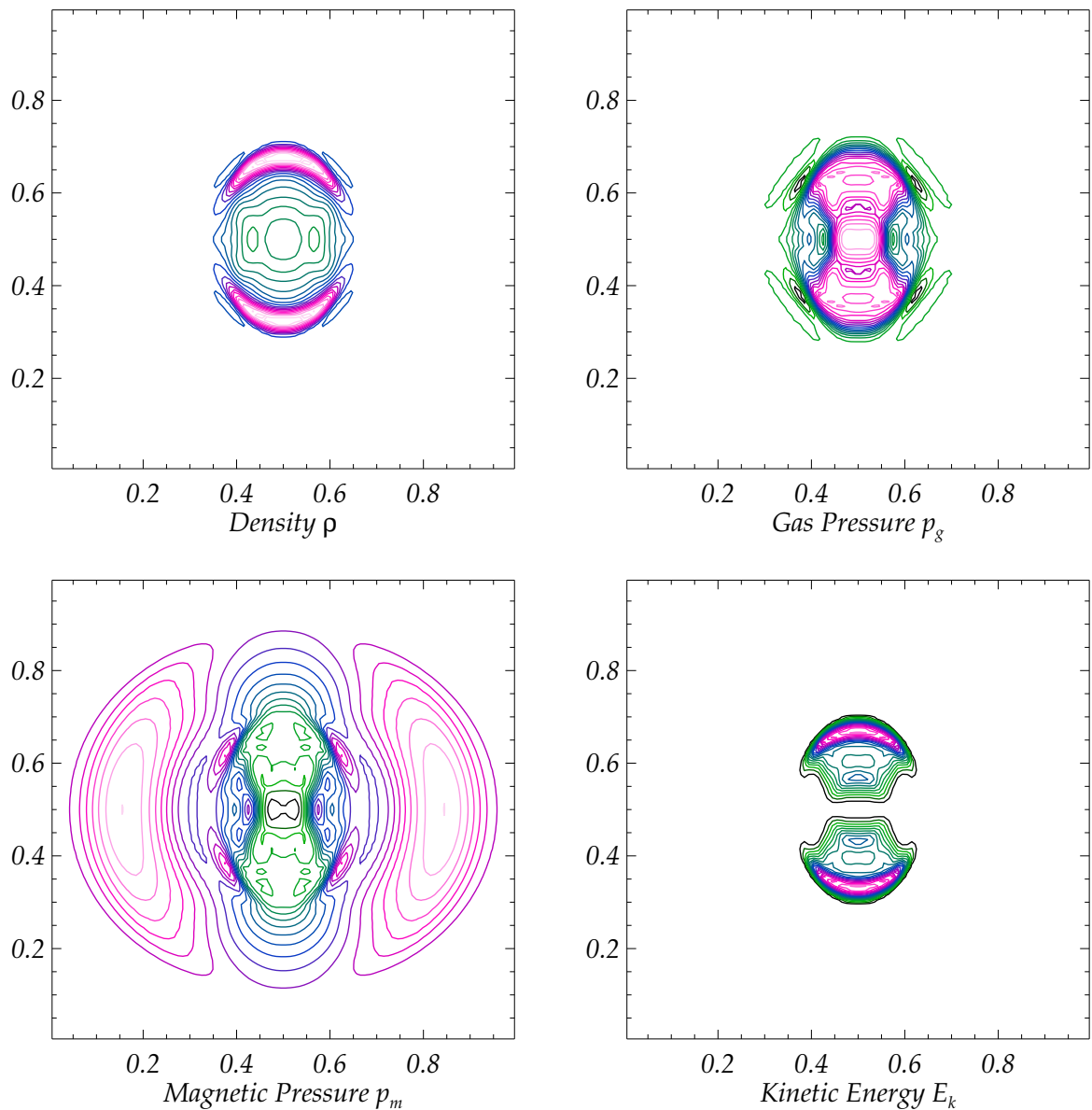}
\caption{A snapshot of the solution to the spherical explosion problem
at time $t=0.0105$. The calculation is performed on a uniform mesh
of 100 $\times$ 100 grids, using the current BGK MHD scheme. 25
contours for each quantity are plotted. Numerical time step is fixed
at $\Delta t=0.008$, corresponding to a Courant number $\approx 0.67$.}
\label{figse}
\end{figure}

Fig.~\ref{figse} shows a snapshot of the solution at $t=0.0105$.
Density, gas pressure, magnetic pressure and kinetic energy are
visualized with 25 contours for each quantities. Due to the strong
magnetic field, the spherical explosion are highly anisotropic. The
shocks also form in the magnetic field.
\begin{figure}
\centering
\includegraphics[scale=0.7]{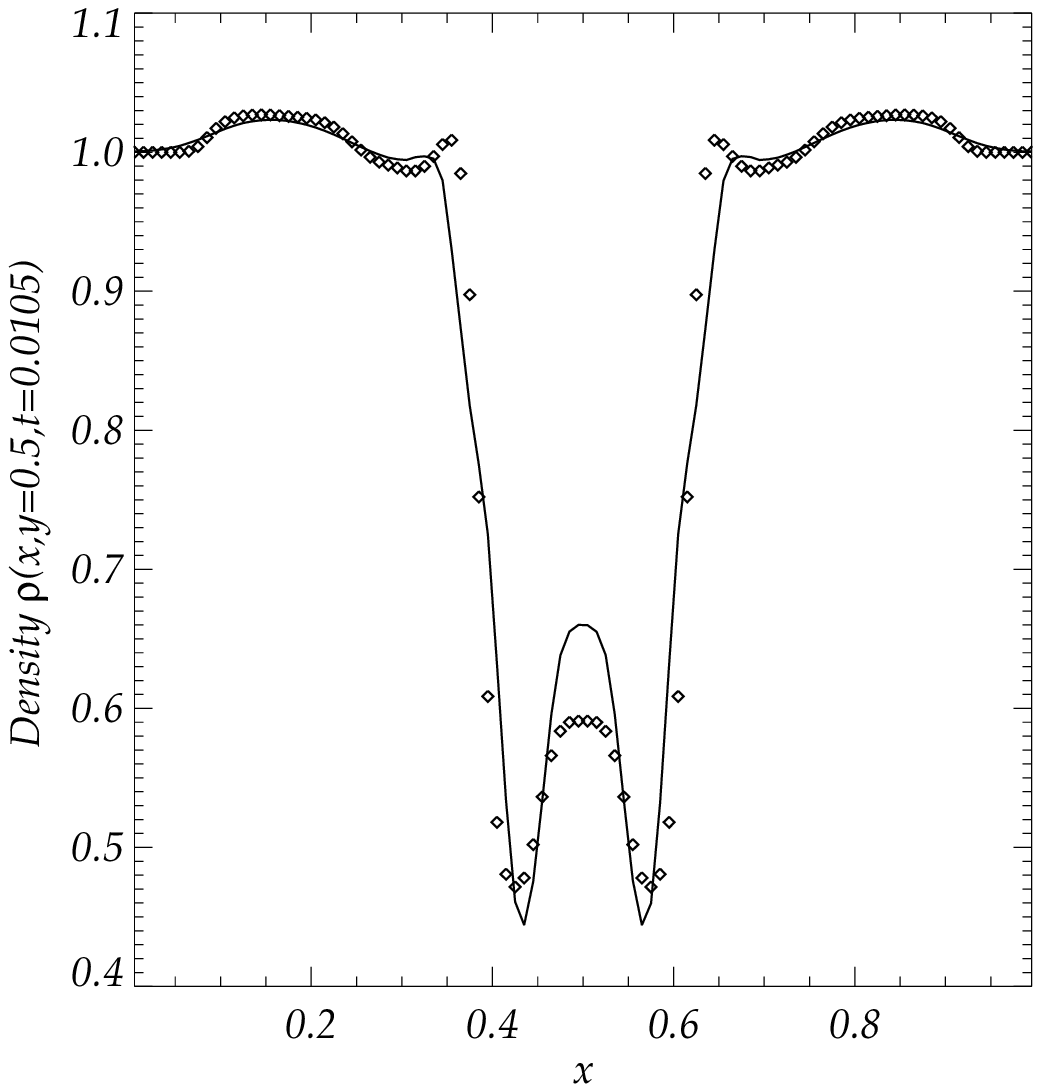}
\caption{The density distributions of the spherical explosion problem
along $y=0.5$ at time $t=0.0105$. Solid line is obtained by the current
BGK MHD scheme. Dots are from the Maxwellian based gas-kinetic scheme
with $\alpha=0.95$.}
\label{figseds}
\end{figure}
Fig.~\ref{figseds} compares the density distributions from
different scheme.  These distributions are taken along $y=0.5$ at
time $t=0.0.105$. Solid line is obtained by the current BGK MHD
scheme. Dots are from Maxwellian based gas-kinetic scheme with
$\alpha=0.95$. The post-shock region indicates that the current
scheme is less dissipative than the Maxwellian based scheme.

\subsection{Orszag-Tang Turbulent Vortex}
\begin{figure}
\centering
\includegraphics[scale=0.9]{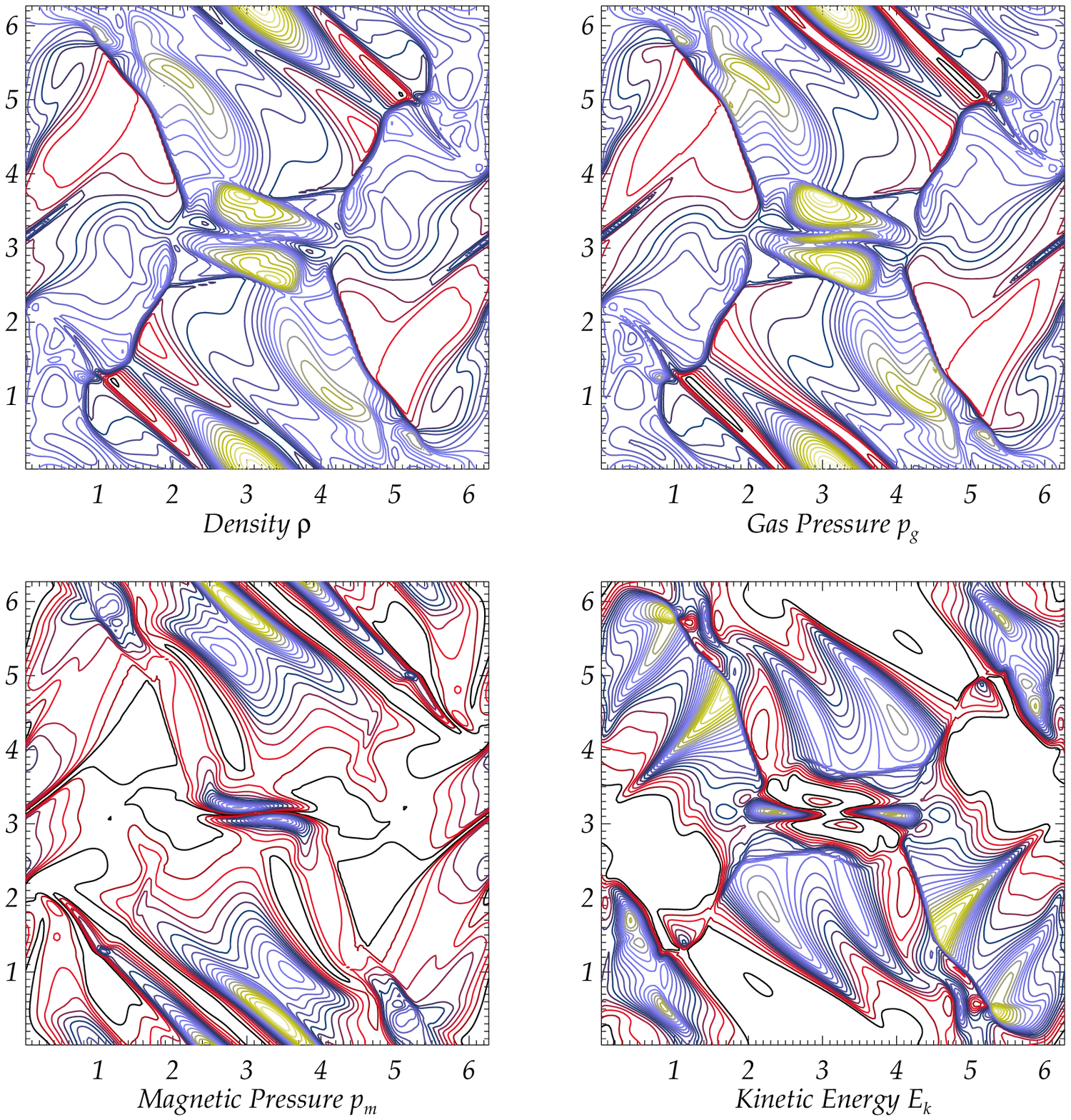}
\caption{A snapshot of the solution to the Orszag-Tang MHD turbulent
problem at time $t=3$. The calculation is performed on a uniform
mesh of 192 $\times$ 192 grids, using BGK MHD scheme without
including the tangential derivatives in the gas evolution stage. 25
contours for each quantity are plotted. Numerical time step is fixed
at $\Delta t=0.008$, corresponding to a Courant number $\approx 0.67$.}
\label{figotv}
\end{figure}

\begin{figure}
\centering
\includegraphics[scale=0.9]{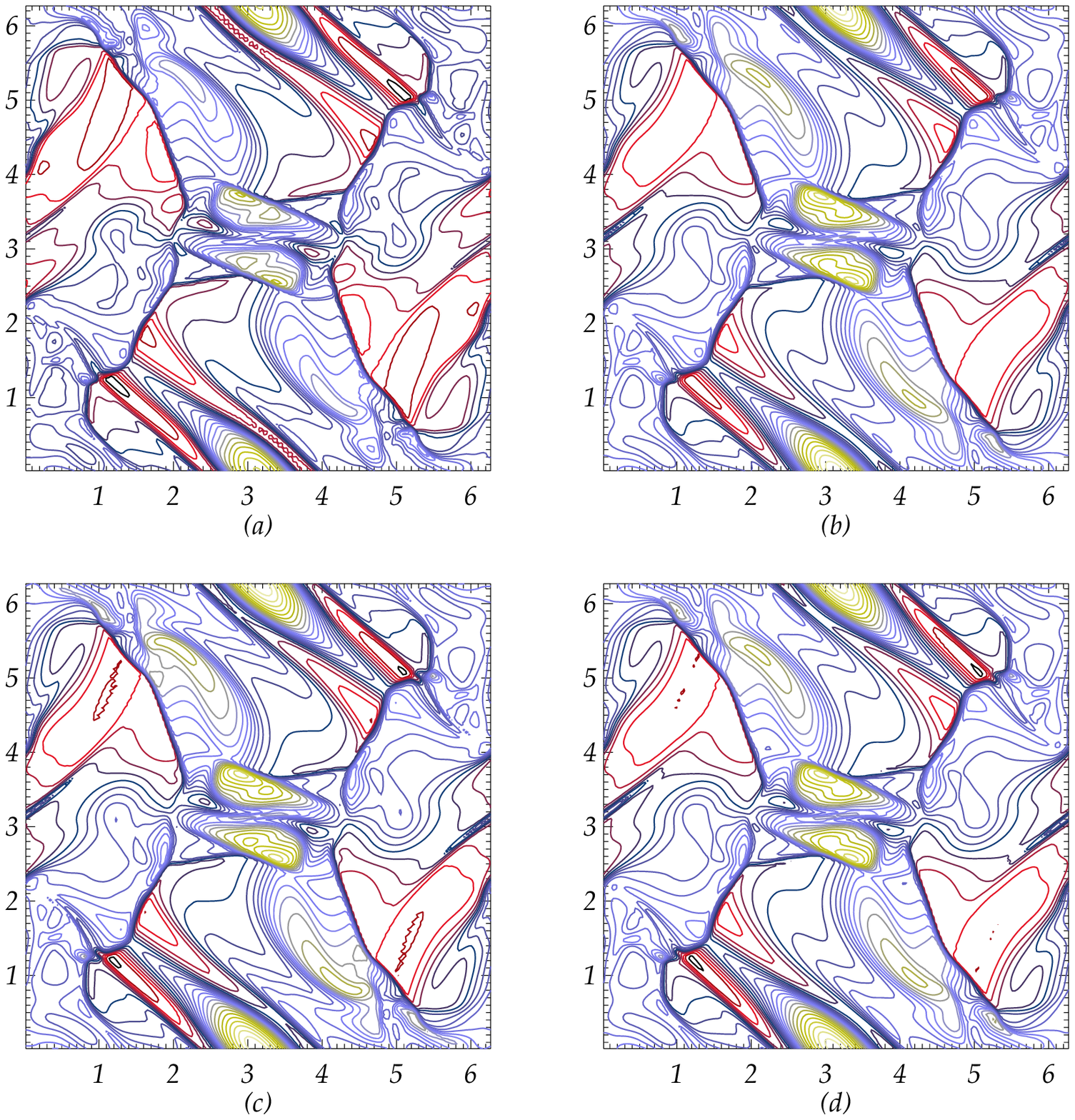}
\caption{The density distributions of the Orszag-Tang problem
at time $t=3$. (a) Maxwellian based gas-kinetic flux
splitting with $\alpha=0.7$; (b) current scheme with Lorentz
force switched off in the gas evolution stage; (c)
tangential derivates switched on; (d) tangential component
of Lorentz force switched on.}
\label{figovtds}
\end{figure}

The 2D Orszag-Tang turbulent vortex problem \cite{orszag79} is
widely used by computational fluid dynamics (CFD) community to test
the MHD scheme because of its complicated interaction between
different waves generated as the vortex system evolving. The
computations are preformed on a domain of $[0:2\pi]\times[0:2\pi]$.
The initial conditions are
\begin{eqnarray}
  \rho(x,y)=\gamma^2, u_x=-\sin{(y)}, u_y=\sin{(x)},\nonumber\\
  p_g(x,y)=\gamma, B_x=-\sin{(y)}, B_y=\sin{(2x)}.
  \label{inicds}
\end{eqnarray}

Fig.~\ref{figotv} shows a snapshot of the solution at time $t=3$.
Twenty five contours for density, gas pressure,magnetic pressure,
kinetic energy are plotted. This solution is obtained by BGK MHD
scheme with a second-order TVD Runge-Kutta time marching scheme. The
effects of tangential derivatives are excluded during gas evolution.
When considering the acceleration due to Lorentz force, only the
normal component is included in the particle distribution function.
The divergence free condition of magnetic field is ensured by
projection method. Around this time, the four spiral arms like shock
fronts are propagating anticlockwise.

Fig.~\ref{figovtds} compares the density distributions
from different splitting methods. A solution of Maxwellian
based gas-kinetic splitting scheme is also displayed in
panel (a). In (b), the Lorzent force is switched off in gas
evolution stage. In (c), the tangential deviations are switched on.
In (d), the tangential component of Lorentz force is included.
The oscillations caused by tangential deviations are evident.
The effect of Lorentz force is not obvious in (b).

\begin{figure}
\centering
\includegraphics[scale=0.9]{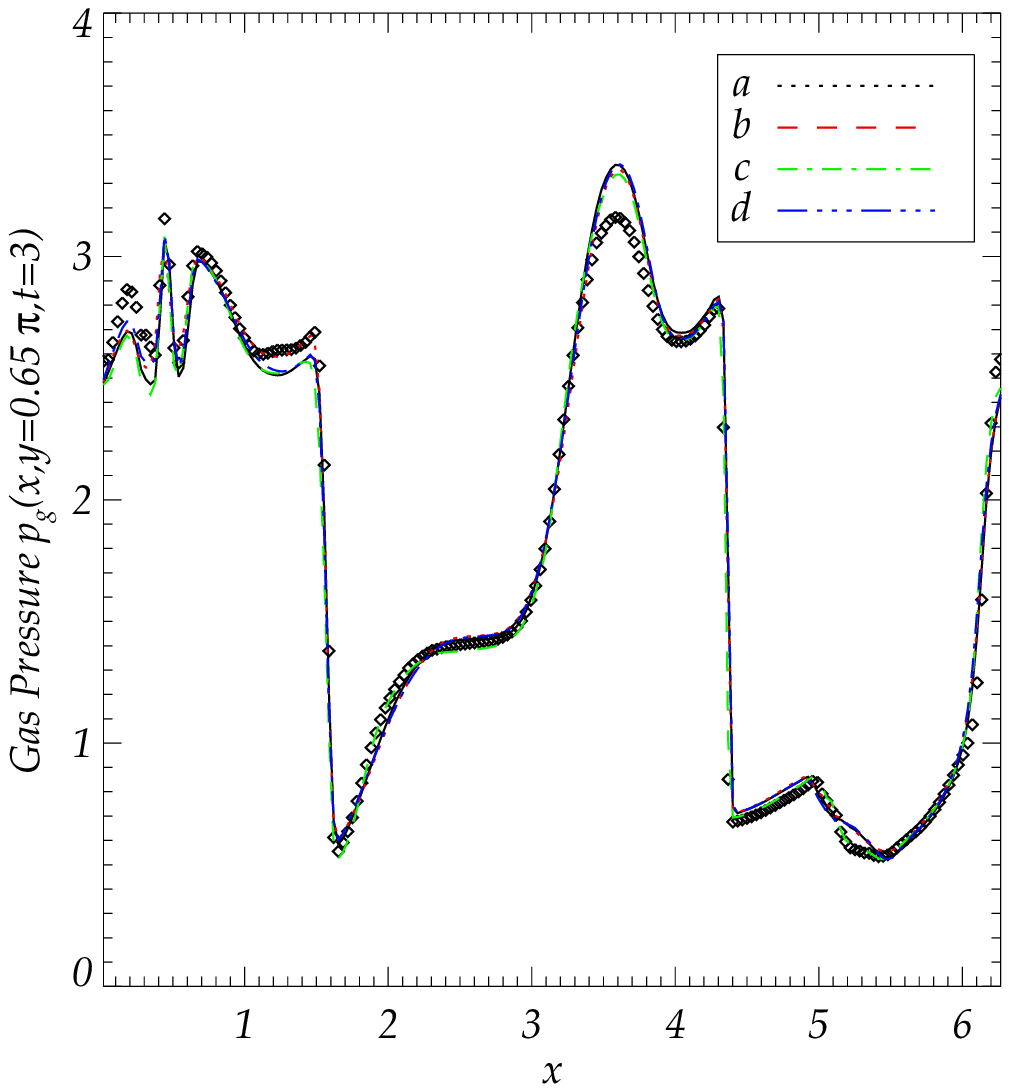}
\caption{Gas pressure distributions along $y=0.625\pi$ at $t=3$.
Dots are from the Maxwellian based gas kinetic flux splitting
scheme. Lines are from BGK MHD scheme. (a)tangential
deviations switched off; (b)Lorentz force switched off in
the gas evolution; (c)tangential derivatives switched on;
(d) tangential component of Lorentz force switched on.}
\label{figovtps}
\end{figure}

A more detailed comparison is given in Fig.~\ref{figovtps}, in which
the gas pressure distributions along $y=0.625\pi$ at $t=3$ are
plotted. At this time, these waves are propagating forward along x
axis. There are three obvious differences in these distributions.
The most obvious different is the altitude of the hump post first
shock -- the shock at $x\sim 4.3$. This hump goes through a long
term evolving. This difference is caused by particle distributions.
Fig.~\ref{figovtps} indicates that the BGK solution based scheme
is less dissipative than the Maxwellian based scheme. The difference
in the fluctuations behind second shock (the one locates at $x\sim
1.6$) is mainly caused by including the Lorentz force effects in the
gas evolution. The fluctuations obtained by current scheme is more
smooth. The third apparent difference occurring at $x\sim 5.25$ is
caused by the implementation of tangential deviations in particle
distribution functions.

\subsection{Cloud-Shock Interaction\cite{dai1998331}}
This simplified astrophysical problem describes the
disruption of a denser cloud by a magnetosonic shock. The
computational domain is $1\times 1$ box solved on a
$200\times 200$ mesh. Initially, the rightward-propagating
shock locates at $x=0.5$ with a Mach number 10. The left and
right states are $(\rho, v_x, v_y, v_z, p_g, B_x, B_y, B_z)
=(3.86859,11.2536,0,0,167.345,0,2.1826182,-2.1826182)$ and
$(1,0,0,0,1,0,0.56418985,0.56418985)$, respectively. The
circular cloud is centered at $(0.25,0.5)$ with a radius
$0.15$ and density $10$. It is in hydrostatic with its
surrounding.

\begin{figure}
\centering
\includegraphics[scale=0.9]{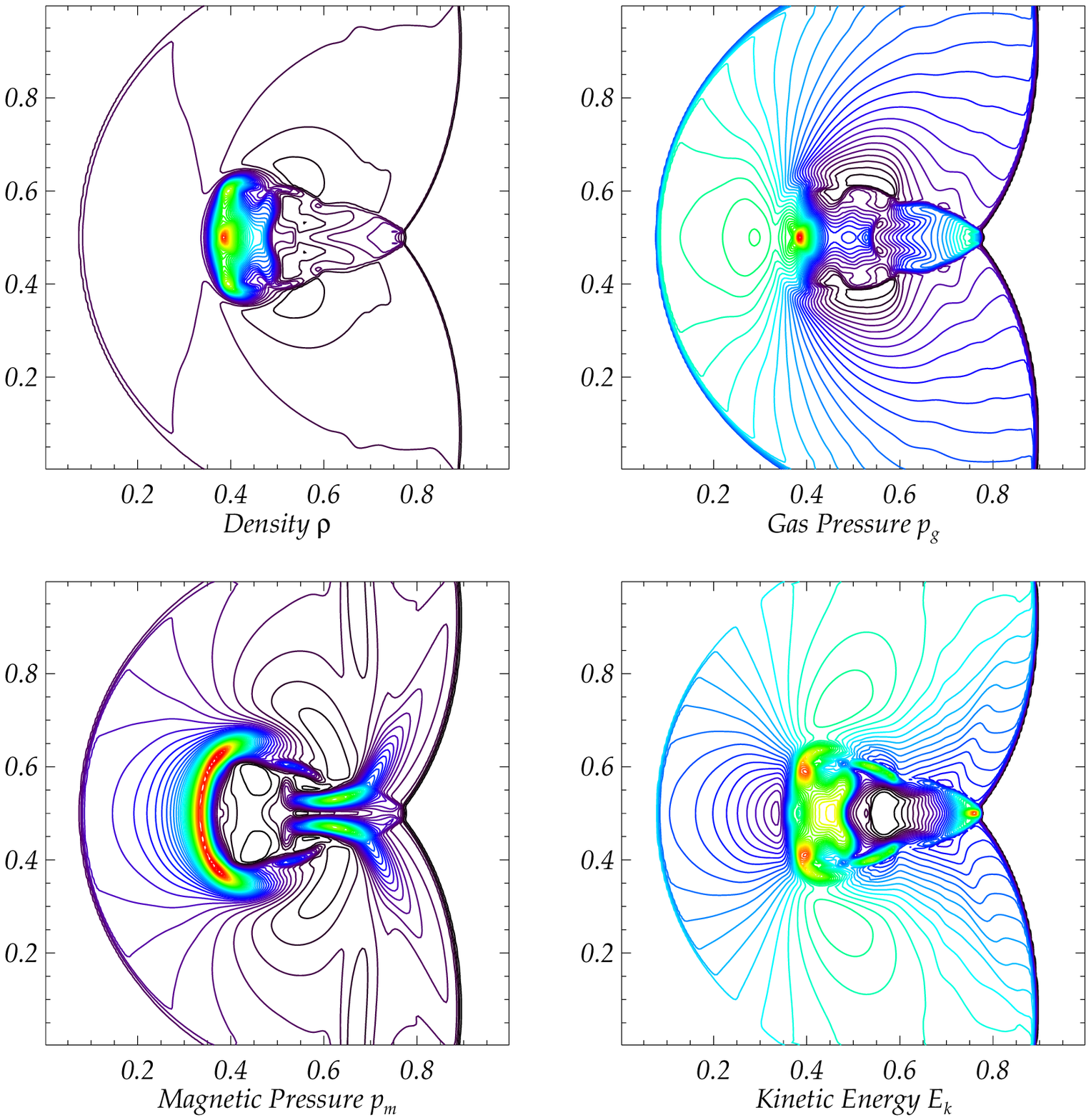}
\caption{The solution of cloud-shock interaction problem
 at time $t=0.055$. The solution is obtained by 3D BGK MHD
 code with a very thin third dimension. Density, gas pressure
 magnetic pressure and kinetic energy are plotted.
 $\nabla\cdot\vec{B}=0$ is ensured by the field-interpolated
 constrained transport method.
 A $200\times 200$ mesh and Curant number $0.7$ are used.}
\label{figssc}
\end{figure}
\begin{figure}
\centering
\includegraphics[scale=0.9]{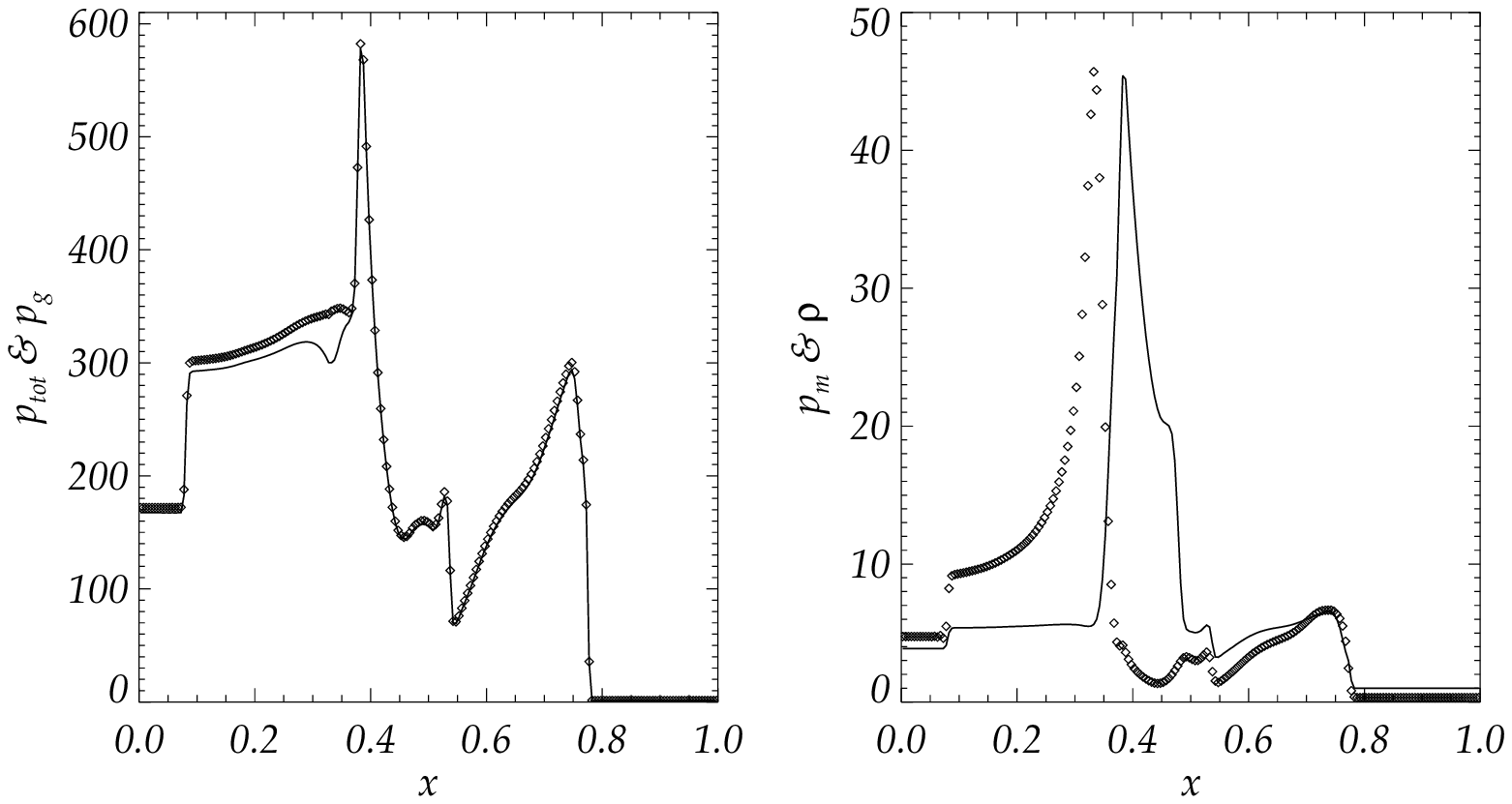}
\caption{The distributions of various quantities  along $y=0.5$
 for cloud-shock interaction problem at time $t=0.055$.
 In the left panel: dots are tot pressure; solid line is
 gas pressure. In the right panel: dots are magnetic pressure;
 solid line is density.}
\label{figscd}
\end{figure}

This problem is solved by our 3D BGK MHD code with a very thin third
dimension. The divergence free condition is ensured by the
field-interpolated constrained transport scheme. The Euler scheme is
used for time-stepping. If the magnetic resistivity is set $\eta=0$,
the calculation easily crashes. Instead of employing the 2nd order
TVD Runge-Kutta scheme and the time-consuming projection method
ensuing $\nabla\cdot\vec{B}=0$, I increase the resistivity
slightly. Numerical experiments show a $\eta=0.008\Delta t$ and
$C_1=0.2$ are enough to keep the computation stable. The tangential
deviations in the distribution function plays a very important role
for this case. If we switch off those tangential deviations, the
oscillations occur widely and crash the computation.
Fig.~\ref{figscd} shows the distributions of various quantities
along the middle line $y=0.5$. It is clear that the current scheme
can capture shock with $4\sim 7$ cells, even with the artificially
enhanced viscosity and resistivity.

\subsection{3D Turbulent Magneto-convection}

\begin{figure}
\centering
\includegraphics[scale=0.75]{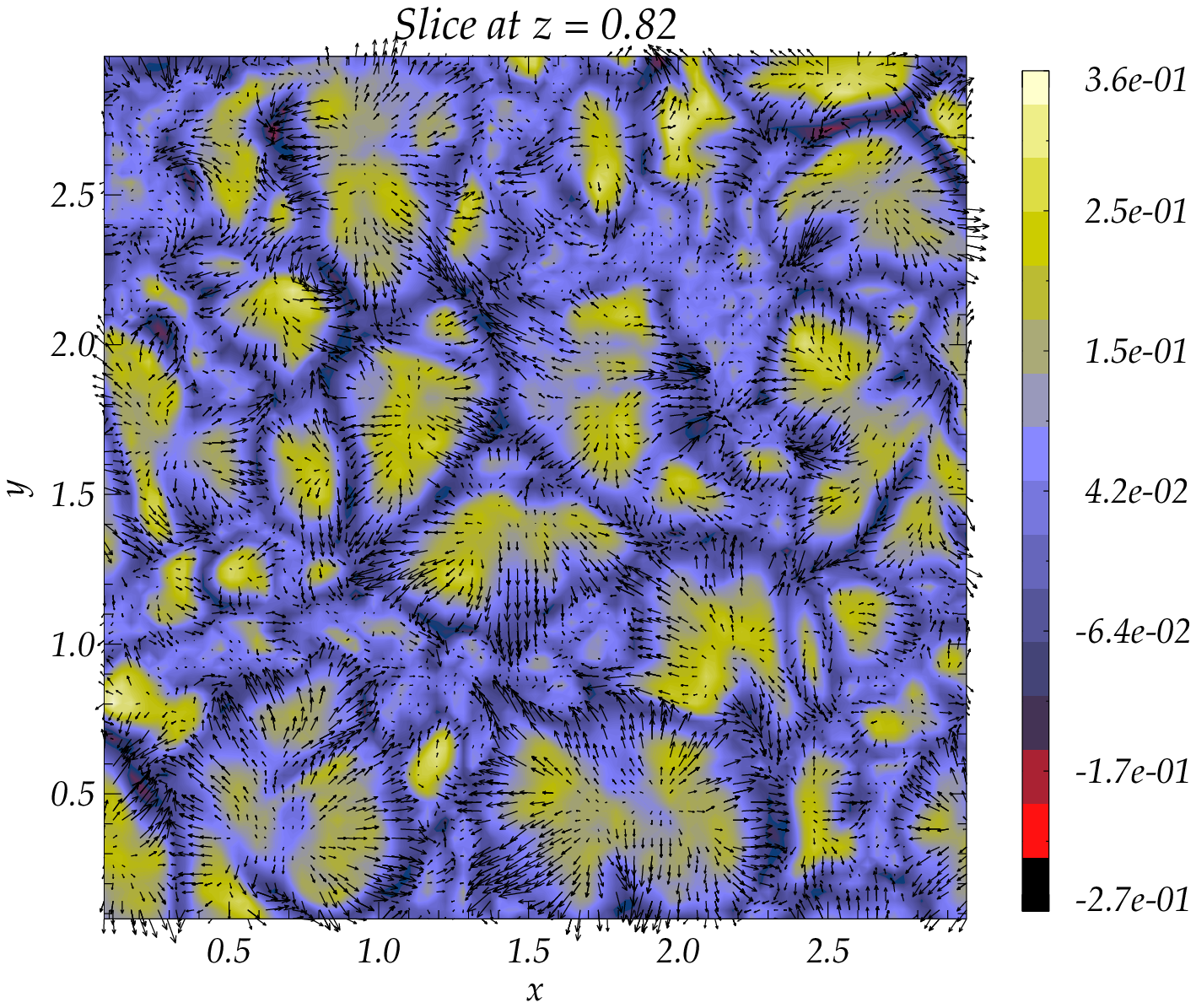}
\includegraphics[scale=0.75]{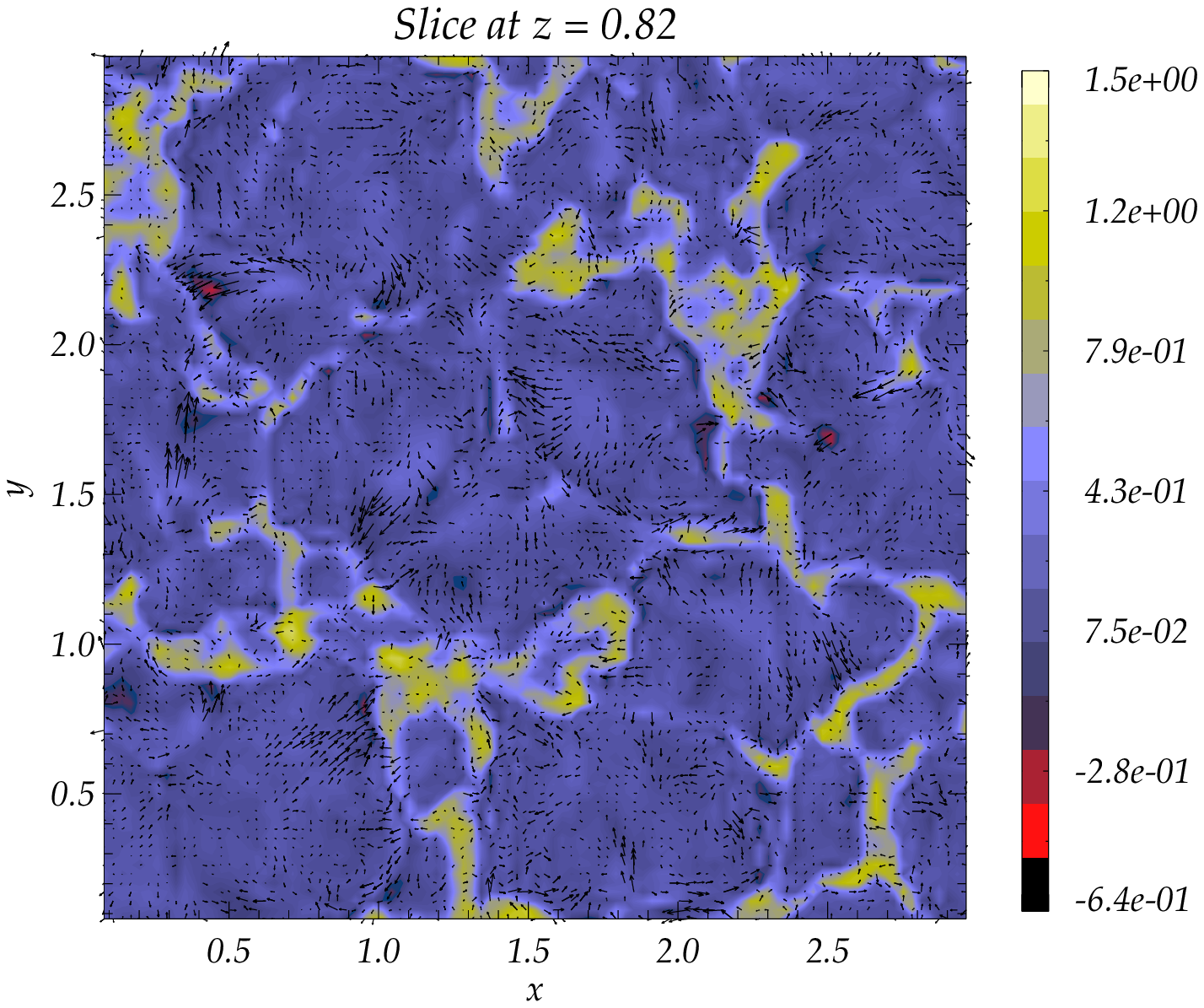}
\caption{Results of 3D turbulent magneto-convection. Horizontal
slices at $z=0.82$. Upper panel: color contours are the vertical
component of the velocity; arrows are the horizontal
components. Lower panel: color contours are the
magnetic strength; arrows are the horizontal components of the
magnetic field.}
\label{fig3dch}
\end{figure}

\begin{figure}
\centering
\includegraphics[scale=0.9]{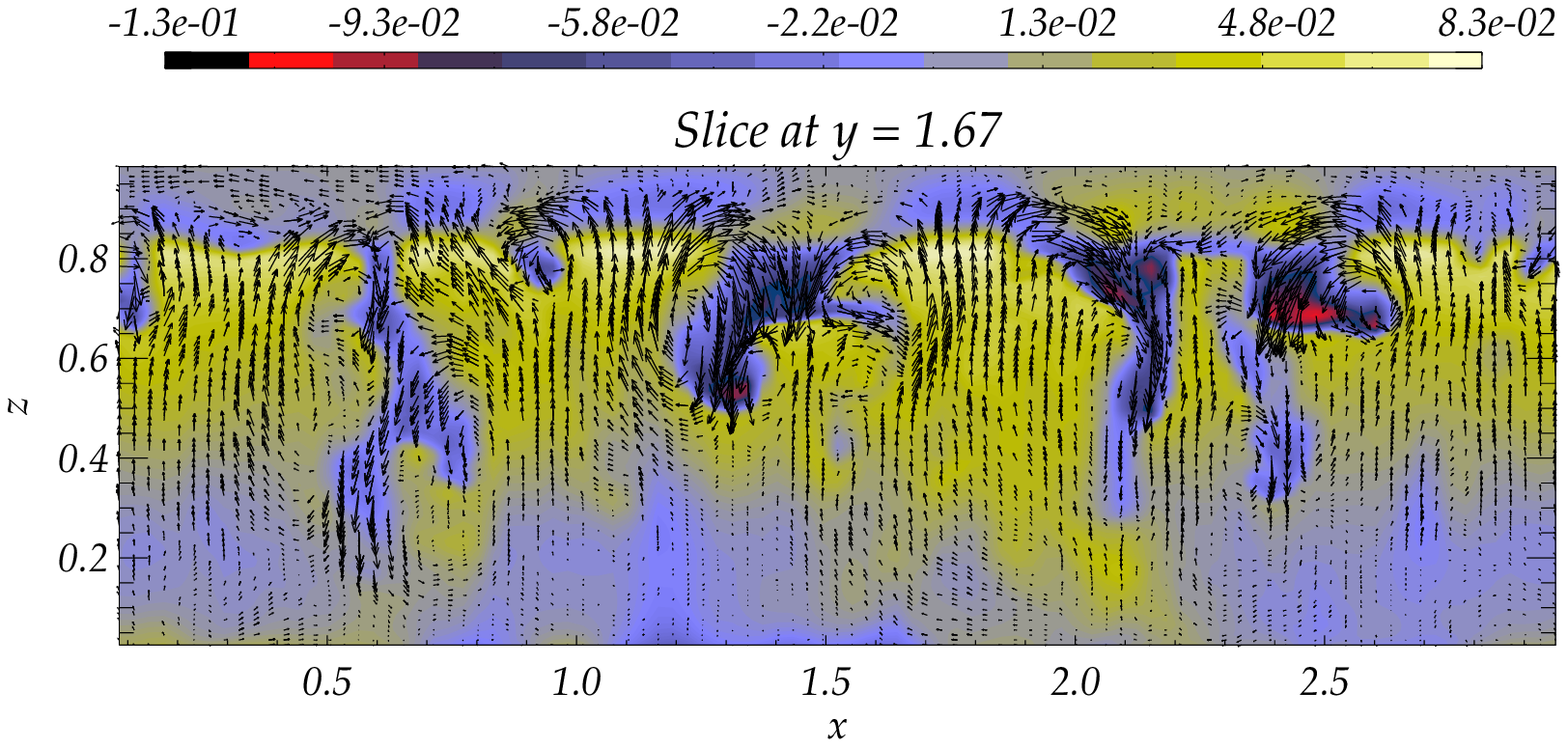}
\includegraphics[scale=0.9]{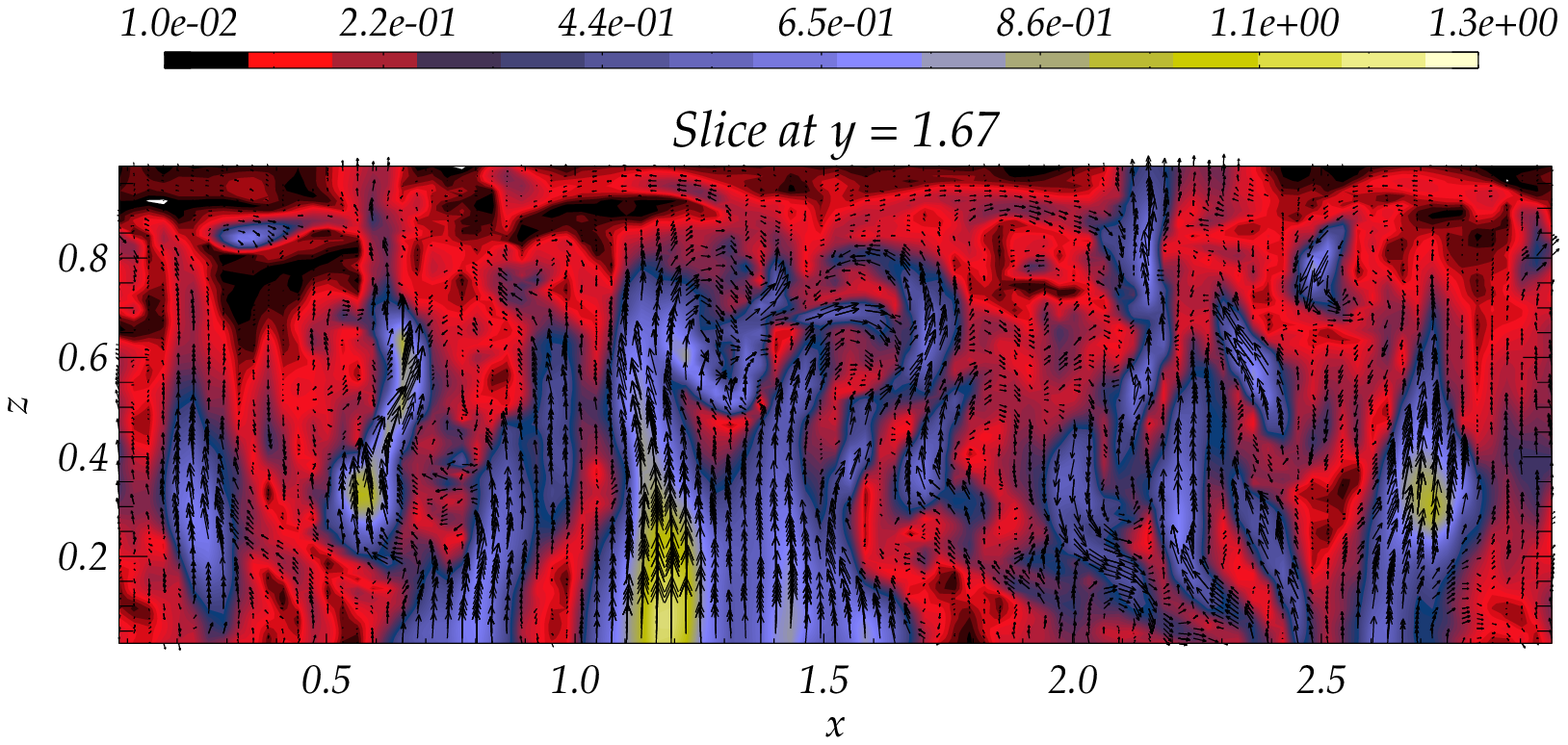}
\caption{Results of 3D turbulent magneto-convection. Vertical slices
at $y=1.67$. Upper panel: color contours are the temperature
fluctuation from horizontal mean; arrows are the projection of the
velocity on x-z plane. Lower panel: color contours are the magnetic
strength; arrows are the projection of the magnetic field on x-z
plane.} \label{fig3dcv}
\end{figure}

\begin{figure}
\centering
\includegraphics[scale=1]{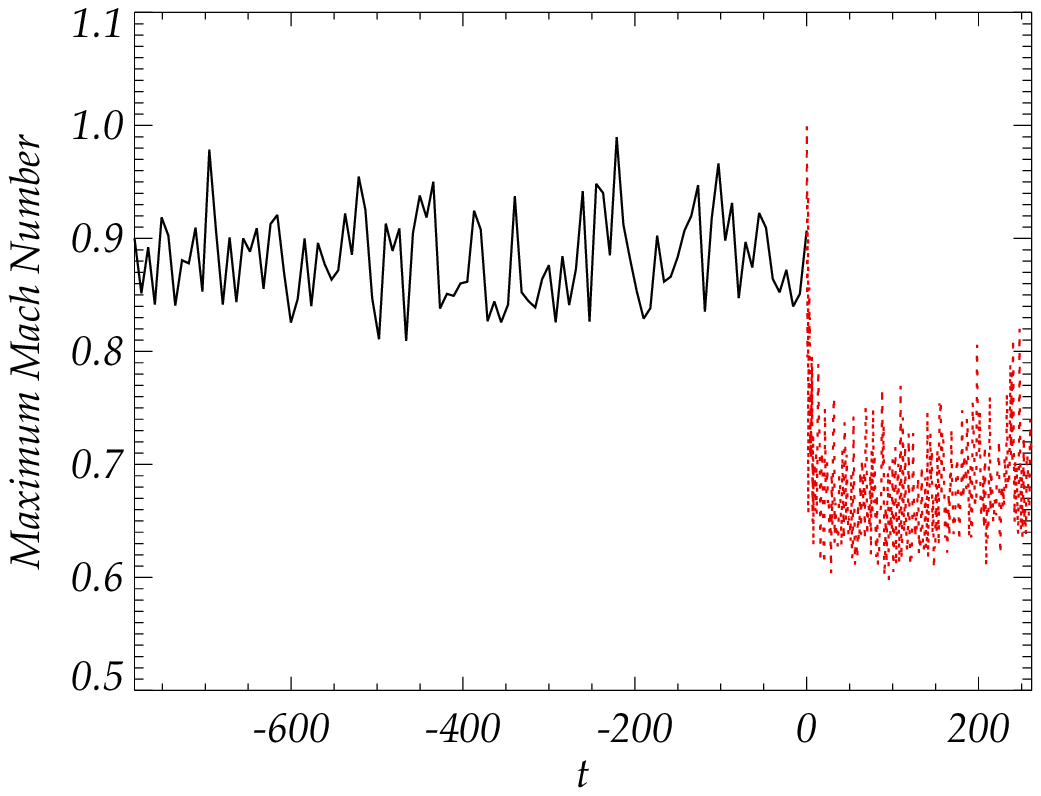}
\caption{Results of 3D turbulent magneto-convection.
Time history of maximum Mach number. $t=0$ is the
initialization time of magentic field. Solid line
indicates the property of non-magnetic convection.
Dotted line indicates the magnetic convection} \label{fig3dmach}
\end{figure}
The 3D turbulent magneto-convection in the stellar interior is
calculated to show the applicability of the current BGK MHD scheme
to complicated astrophysical flows. This is one of the most
difficult MHD problem in astrophysics. It is essentially a multi-
length-scale and time-scale problem, which makes its numerical
modeling very time- and memory- consuming. In the deep stellar
atmosphere, the convective motions are subsonic. However just
beneath the upper radiative zone, there is thin highly
superadiabatic layer, where complex waves, including supersonic
waves generated. The computation is easily interrupted if the code
is not robust, especially, during the thermal relaxation phase of
the model.

Here I calculate a 3D sandwich model of stellar convection:
one convective layer locates in between two stable layers.
The stable layers are radiative. Radiation transfer is
treated by diffusion approximation. Under constant gravity,
the system is initially in hydrostatic state and vertically
stratified. The method to construct the initial
stratification can be find in \cite{src95}. The middle layer
is slightly superadiabatic. The stable layers are
subadiabatic. The vertical depth extents $\sim 5.3$ pressure scale
heights. The sub-grid scale turbulent behaviors are mimicked
by Smagorinsky model\cite{sgs63,tian09a}. The aspect ratio
is $3/1$ (width/depth). The horizontal boundaries are
periodic. The upper boundary and lower boundary are
impenetrable. All quantities are scaled such that the
density, pressure, temperature at the top boundary are unit.
The length is scaled by the depth of the computation domain.
This initial state will undergo an adjustment, significantly
near the vertical boundaries and stable-unstable interface,
moderately in the convective region. Finally, the system
will be thermally relaxed and approach an dynamic
equilibrium state, indicated by the balance between outgoing
energy flux through top and input energy flux from bottom.

When the non-magneto convection nearly approaches the
relaxed state, I impose a uniform vertical magnetic field
to a snapshot of the numerical solutions. Its strength
equals the equipartition value near the top. At the top and
bottom, the vertical field boundary condition is adopted for
the magnetic field, i.e.,
\begin{equation}
  B_x=B_y=0, \frac{\partial B_z}{\partial z}=0.
  \label{bbcs}
\end{equation}
The computation was run on the supercomputer with a 336
threading code parallelized by massage passing interface
(MPI) protocol. The computation of non-magnetic convection
took $\sim 36$ days, it is still not completely relaxed. The
upper boundary can transport around 85 percent of the input
energy fluxes. The magneto convection was run for $\sim 7$
days. At the end, 93 percent input fluxes can be transported
out from upper lid.

Fig.~\ref{fig3dch} shows the projections of 3D velocity and
magnetic field on a horizontal x-y plane, at $z=0.82$. Upper panel
shows the velocity field. Lower panel shows the magnetic field.
Color contours are the vertical component. Arrows are horizontal
fields. Fig.~\ref{fig3dcv} displays vertical slices of the flows.
In upper panel, the 2D velocity field is imposed on the fluctuation
of temperature from the horizontal mean. In lower panel, the 2D
magnetic field is imposed on the contour of magnetic strength.
Fig.~\ref{fig3dmach} shows the time history of the maximum Mach
number of the system. The magnetic field suppresses the motion
evidently. Before superposing magnetic field, the highest speed
motion is nearly supersonic. 

\section{Discussion}
\label{discu}

The current scheme can be regarded as a hybrid method, a combination
of BGK-NS solver and flux splitting scheme. Without introducing
parameter-controlled equilibrium state, the treatment of the
magnetic part of fluxes is similar to the FVS scheme. Unlike the FVS
scheme and the gas-kinetic MHD method developed by
Xu\cite{xu1999334}, where the Maxwellian distribution is used, the
current method splits the solution of BGK equation. The collision
mechanism is already included in the split distribution. Physically,
the current scheme is supposed to be better. However, numerical
tests show that the improvement is not very great. For some cases,
it is dispersive.

For non-magnetic flows, the tangential deviations from equilibrium
state in particle distribution function play a role in improving the
accuracy of the scheme. After implementing the magnetic field, the
effects become case-dependent. The effects of tangential terms can
be thought as reducing the numerical dissipation. For some cases,
eg., 2D Orzang-Tang vortex, this kind of reduction seems too much,
even with a 2nd order TVD Runge-Kutta time marching scheme. To get
around with this kind of situation, it is better to switch off these
tangential deviations terms in particle distribution functions.

As stated before, the particle solution from solving BGK
equation can be split into three parts, i.e., $[\cdots]g_0$,
split parts $[\cdots]g_0^l$ and $[\cdots]g_0^r$, just like
Maxwellian based scheme, i.e., $(1-\alpha) g_0$, $\alpha
g_0^l$ and $\alpha g_0^r$. The adjustable parameter is
replaced by complicated coefficients from solving BGK
equation. It sounds more physical. Numerical tests again
show that this kind of consideration would make a very
dispersive scheme. This is because in Maxwellian based
scheme, the particle collisions are controlled by an
universal constant $\alpha$, so the numerical dissipation
from MUSCL-type reconstruction is added in whole computational
domain. In the BGK scheme, the numerical dissipation only
become large near the region where the particle distribution
deviates from equilibrium state significantly. More
specifically, the equilibrium term and non-equilibrium terms
from (\ref{stbgk}) can be expressed as
$[1+e^{-t/\tau}+\cdots]g_0$, split parts
$[e^{-t/\tau}(\cdots)]g_0^l$ and
$[e^{-t/\tau}(\cdots)]g_0^r$. $e^{-t/\tau}$ is a very small
number, because usually we have $\tau\lesssim t/10$, which
means the split parts can be ignored. $\tau$ only becomes
comparable to $t$ near the high non-equilibrium region. This
is reason why BGK is a very robust and accurate scheme for
non-magnetic shock wave problems. For magnetic flows, the
equilibrium part $[1+e^{-t/\tau}+\cdots]g_0$ needs to be
split also to include enough dissipation to smooth the
dispersion.

Fig.~\ref{figcop2} shows clearly that the dispersion
occurs near the front of fast wave and propagates backward.
Microscopically, transport of flow properties can be
regarded as a consequence of particle motions and the amount
of flux is proportional to the number of involved
particles. This is not true for the magnetic field, even
those advective terms. This is because the unlike charge or
mass, there is no specific amount magnetic is carried by
each particle. If we treat the non-magnetic part by a second
order time scheme, while the magnetic field by a first oder
time scheme. Although, the magnetic terms are weighted by
second order particle distributions, the difference of time
accuracy of these two parts is still big. The truncation
errors propagate at different speed, which amplifies the
numerical dispersion. A second-order TVD Runge-Kutta scheme
can smear out these oscillations.

The second-order TVD time scheme usually catches a sloping
shock front. A way to improve this is to design a very smart
parameter $\alpha$ that can introduce numerical dissipation
both near the shock and the fast waves. In \cite{Tang20103}
  $\alpha$ is adaptive according to the discontinuity in
magnetic pressure. A more consistent way to construct a
gas-kinetic theory based MHD scheme is to considers
separately the distribution of charged particle (electrons
and ions) and electrically neutral particle. Model the long
distance interactions between charged particles and then
integrate the electric field and magnetic properties from
the motion of these charged particles. In such case, the EOS
play a very important role, since it determines the number
of charged particles. Obviously, this is a very complicated
method. A more detailed discussion is beyond the scope of
current study.

The gas-kinetic MHD scheme developed in \cite{Tang200069} is
an efficient method. For the 2D Orzang-Tang vortex problem
tested in this paper, it can be two times faster than the
current scheme (with $\sim 0.47$ time-consuming). However,
in application to astrophysical problems, a lot of things,
such as radiation transfer, turbulent modeling, moving mesh
generating, need to be implemented. The hydrodynamics part
of the code is not the major agent consuming the computation
time. Sometimes, even the ensuring divergence free constraint
takes a lot time. For instance, for the 2D tests in this paper, a
Possion equation is solved to ensue $\nabla\cdot\vec{B}=0$.
This involves an implicit solver. The time-consuming of an
implicit solver is generally not linearly proportional to
the grid points. The MHD code based on current scheme is
capable of calculating the complicated HD problem without
extra efforts. At the same time, it is also a high
order MHD solver.

\section{Conclusion}
\label{conclu}

This paper presents a feasible way to extend the
multidimensional gas-kinetic BGK scheme to MHD problems. The
non-magnetic part is solved by the BGK scheme modified due
to magnetic field. While the magnetic part is calculated by
gas-kinetic flux splitting based on a solution of BGK
equations. Besides keeping high robustness and accuracy for
HD problem, the current scheme is also a high order MHD
solver. The effects of tangential deviations and Lorentz
force in particle distribution are numerically studied.
Although case-dependent, it is better to include them
for most of the cases.  Numerical tests also show
that compared to the Maxwellian based gas-kinetic flux
splitting scheme, the current scheme is more stable for 1D
problems and less dissipative for multidimensional problems.
The 3D turbulent magneto-convection test indicates the
applicability of current scheme to complicated astrophysical
flows. The gas-kinetic theory based flux splitting methods
for MHD problems can be improved by designing a more smart
parameter $\alpha$. A full gas-kinetic scheme for MHD needs
to consider the complicated physical processes, such as
particle charging and long distance Coulomb force.

\section{Acknowledgments}
The computations of 3D turbulent magneto-convection were
tested and performed on E\"{o}v\"{o}s University's 416-core,
3.7 Tflop HPC cluster {\it Atlasz} and on the supercomputing
network of the NIIF Hungarian National Supercomputing Center
(project ID: 1117 fragment). The current research was
supported by the European Commission's 6th Framework
Programme (SOLAIRE Network, MTRN-CT-2006-035484).

\bibliographystyle{model1b-num-names}
\bibliography{tian,xu,num,chan}

\begin{thebibliography}{22}
\expandafter\ifx\csname natexlab\endcsname\relax\def\natexlab#1{#1}\fi
\providecommand{\bibinfo}[2]{#2}
\ifx\xfnm\relax \def\xfnm[#1]{\unskip,\space#1}\fi
\bibitem[{Bhatnagar et~al.(1954)Bhatnagar, Gross and Krook}]{bgk54}
\bibinfo{author}{P.L. Bhatnagar}, \bibinfo{author}{E.P. Gross},
  \bibinfo{author}{M.~Krook}, \bibinfo{title}{A model for collision processes
  in gases. i. small amplitude processes in charged and neutral one-component
  systems}, \bibinfo{journal}{Phys. Rev.} \bibinfo{volume}{94}
  (\bibinfo{year}{1954}) \bibinfo{pages}{511--525}.
\bibitem[{Brio and Wu(1988)}]{Brio1988400}
\bibinfo{author}{M.~Brio}, \bibinfo{author}{C.C. Wu}, \bibinfo{title}{An upwind
  differencing scheme for the equations of ideal magnetohydrodynamics},
  \bibinfo{journal}{Journal of Computational Physics} \bibinfo{volume}{75}
  (\bibinfo{year}{1988}) \bibinfo{pages}{400 -- 422}.
\bibitem[{Colella and Woodward(1984)}]{ppm1984}
\bibinfo{author}{P.~Colella}, \bibinfo{author}{P.R. Woodward},
  \bibinfo{title}{The piecewise parabolic method (ppm) for gas-dynamical
  simulations}, \bibinfo{journal}{Journal of Computational Physics}
  \bibinfo{volume}{54} (\bibinfo{year}{1984}) \bibinfo{pages}{174 -- 201}.
\bibitem[{Dai and Woodward(1994)}]{Dai1994485}
\bibinfo{author}{W.~Dai}, \bibinfo{author}{P.R. Woodward},
  \bibinfo{title}{Extension of the piecewise parabolic method to
  multidimensional ideal magnetohydrodynamics}, \bibinfo{journal}{Journal of
  Computational Physics} \bibinfo{volume}{115} (\bibinfo{year}{1994})
  \bibinfo{pages}{485 -- 514}.
\bibitem[{Dai and Woodward(1998)}]{dai1998331}
\bibinfo{author}{W.~Dai}, \bibinfo{author}{P.R. Woodward}, \bibinfo{title}{A
  simple finite difference scheme for multidimensional magnetohydrodynamical
  equations}, \bibinfo{journal}{Journal of Computational Physics}
  \bibinfo{volume}{142} (\bibinfo{year}{1998}) \bibinfo{pages}{331 -- 369}.
\bibitem[{Harten(1983)}]{Harten1983357}
\bibinfo{author}{A.~Harten}, \bibinfo{title}{High resolution schemes for
  hyperbolic conservation laws}, \bibinfo{journal}{Journal of Computational
  Physics} \bibinfo{volume}{49} (\bibinfo{year}{1983}) \bibinfo{pages}{357 --
  393}.
\bibitem[{{Orszag} and {Tang}(1979)}]{orszag79}
\bibinfo{author}{S.A. {Orszag}}, \bibinfo{author}{C.M. {Tang}},
  \bibinfo{title}{{Small-scale structure of two-dimensional magnetohydrodynamic
  turbulence}}, \bibinfo{journal}{Journal of Fluid Mechanics}
  \bibinfo{volume}{90} (\bibinfo{year}{1979}) \bibinfo{pages}{129--143}.
\bibitem[{{Ryu} and {Jones}(1995)}]{ryu951}
\bibinfo{author}{D.~{Ryu}}, \bibinfo{author}{T.W. {Jones}},
  \bibinfo{title}{{Numerical magetohydrodynamics in astrophysics: Algorithm and
  tests for one-dimensional flow`}}, \bibinfo{journal}{\apj}
  \bibinfo{volume}{442} (\bibinfo{year}{1995}) \bibinfo{pages}{228--258}.
\bibitem[{{Ryu} et~al.(1995){Ryu}, {Jones} and {Frank}}]{ryu952}
\bibinfo{author}{D.~{Ryu}}, \bibinfo{author}{T.W. {Jones}},
  \bibinfo{author}{A.~{Frank}}, \bibinfo{title}{{Numerical Magnetohydrodynamics
  in Astrophysics: Algorithm and Tests for Multidimensional Flow}},
  \bibinfo{journal}{\apj} \bibinfo{volume}{452} (\bibinfo{year}{1995})
  \bibinfo{pages}{785--+}.
\bibitem[{{Shu}(1988)}]{shu88}
\bibinfo{author}{C.W. {Shu}}, \bibinfo{title}{{Total-Variation-Diminishing Time
  Discretizations}}, \bibinfo{journal}{SIAM J. Sci. and Stat. Comput.}
  \bibinfo{volume}{9} (\bibinfo{year}{1988}) \bibinfo{pages}{1073--1084}.
\bibitem[{{Singh} et~al.(1995){Singh}, {Roxburgh} and {Chan}}]{src95}
\bibinfo{author}{H.P. {Singh}}, \bibinfo{author}{I.W. {Roxburgh}},
  \bibinfo{author}{K.L. {Chan}}, \bibinfo{title}{{Three-dimensional simulation
  of penetrative convection: penetration below a convection zone.}},
  \bibinfo{journal}{\aap} \bibinfo{volume}{295} (\bibinfo{year}{1995})
  \bibinfo{pages}{703--+}.
\bibitem[{{Smagorinsky}(1963)}]{sgs63}
\bibinfo{author}{J.~{Smagorinsky}}, \bibinfo{title}{{General Circulation
  Experiments with the Primitive Equations}}, \bibinfo{journal}{Monthly Weather
  Review} \bibinfo{volume}{91} (\bibinfo{year}{1963}) \bibinfo{pages}{99--+}.
\bibitem[{Steger and Warming(1981)}]{Steger1981263}
\bibinfo{author}{J.L. Steger}, \bibinfo{author}{R.F. Warming},
  \bibinfo{title}{Flux vector splitting of the inviscid gasdynamic equations
  with application to finite-difference methods}, \bibinfo{journal}{Journal of
  Computational Physics} \bibinfo{volume}{40} (\bibinfo{year}{1981})
  \bibinfo{pages}{263 -- 293}.
\bibitem[{Tang and Xu(2000)}]{Tang200069}
\bibinfo{author}{H.Z. Tang}, \bibinfo{author}{K.~Xu}, \bibinfo{title}{A
  high-order gas-kinetic method for multidimensional ideal
  magnetohydrodynamics}, \bibinfo{journal}{Journal of Computational Physics}
  \bibinfo{volume}{165} (\bibinfo{year}{2000}) \bibinfo{pages}{69 -- 88}.
\bibitem[{Tang et~al.(2010)Tang, Xu and Cai}]{Tang20103}
\bibinfo{author}{H.Z. Tang}, \bibinfo{author}{K.~Xu}, \bibinfo{author}{C.~Cai},
  \bibinfo{title}{Gas-kinetic bgk scheme for three dimensional
  magnetohydrodynamics}, \bibinfo{journal}{Numer. Math. Theor. Meth. Appl.}
  \bibinfo{volume}{3} (\bibinfo{year}{2010}) \bibinfo{pages}{387 -- 404}.
\bibitem[{{Tian} et~al.(2009){Tian}, {Deng}, {Chan} and {Xiong}}]{tian09a}
\bibinfo{author}{C.~{Tian}}, \bibinfo{author}{L.~{Deng}},
  \bibinfo{author}{K.~{Chan}}, \bibinfo{author}{D.~{Xiong}},
  \bibinfo{title}{{Efficient turbulent compressible convection in the deep
  stellar atmosphere}}, \bibinfo{journal}{Research in Astronomy and
  Astrophysics} \bibinfo{volume}{9} (\bibinfo{year}{2009})
  \bibinfo{pages}{102--114}.
\bibitem[{{Tian} et~al.(2007){Tian}, {Xu}, {Chan} and {Deng}}]{tian07}
\bibinfo{author}{C.L. {Tian}}, \bibinfo{author}{K.~{Xu}}, \bibinfo{author}{K.L.
  {Chan}}, \bibinfo{author}{L.C. {Deng}}, \bibinfo{title}{{A three-dimensional
  multidimensional gas-kinetic scheme for the Navier Stokes equations under
  gravitational fields}}, \bibinfo{journal}{Journal of Computational Physics}
  \bibinfo{volume}{226} (\bibinfo{year}{2007}) \bibinfo{pages}{2003--2027}.
\bibitem[{T\'{o}th(2000)}]{toth2000605}
\bibinfo{author}{G.~T\'{o}th}, \bibinfo{title}{The $\nabla\cdot\vec{B}=0$
  constraint in shock-capturing magnetohydrodynamics codes},
  \bibinfo{journal}{Journal of Computational Physics} \bibinfo{volume}{161}
  (\bibinfo{year}{2000}) \bibinfo{pages}{605 -- 652}.
\bibitem[{Xu(1999)}]{xu1999334}
\bibinfo{author}{K.~Xu}, \bibinfo{title}{Gas-kinetic theory-based flux
  splitting method for ideal magnetohydrodynamics}, \bibinfo{journal}{Journal
  of Computational Physics} \bibinfo{volume}{153} (\bibinfo{year}{1999})
  \bibinfo{pages}{334 -- 352}.
\bibitem[{Xu(2001)}]{xu2001289}
\bibinfo{author}{K.~Xu}, \bibinfo{title}{A gas-kinetic bgk scheme for the
  navier-stokes equations and its connection with artificial dissipation and
  godunov method}, \bibinfo{journal}{Journal of Computational Physics}
  \bibinfo{volume}{171} (\bibinfo{year}{2001}) \bibinfo{pages}{289 -- 335}.
\bibitem[{Xu and Prendergast(1994)}]{Xu19949}
\bibinfo{author}{K.~Xu}, \bibinfo{author}{K.H. Prendergast},
  \bibinfo{title}{Numerical navier-stokes solutions from gas kinetic theory},
  \bibinfo{journal}{Journal of Computational Physics} \bibinfo{volume}{114}
  (\bibinfo{year}{1994}) \bibinfo{pages}{9 -- 17}.
\bibitem[{Zachary et~al.(1994)Zachary, Malagoli and Colella}]{zachary263}
\bibinfo{author}{A.L. Zachary}, \bibinfo{author}{A.~Malagoli},
  \bibinfo{author}{P.~Colella}, \bibinfo{title}{A higher-order godunov method
  for multidimensional ideal magnetohydrodynamics}, \bibinfo{journal}{SIAM
  Journal on Scientific Computing} \bibinfo{volume}{15} (\bibinfo{year}{1994})
  \bibinfo{pages}{263--284}.

\end{thebibliography}
\end{document}